\documentclass[traditabstract]{aa}  
\usepackage{graphicx}
\usepackage{xcolor}
\usepackage{rotating}
\usepackage{tikz}
\usepackage{txfonts}
\usepackage{supertabular}
\usepackage{longtable}
\usepackage{tabularx}
\usepackage{lscape}
\usepackage{soul}
\usepackage{appendix}
\usepackage{float}
\usepackage{natbib}
\usepackage{hyperref}
\bibpunct{(}{)}{;}{a}{}{,} 
\usepackage{adjustbox}
\usepackage{subfig}
\usepackage{multicol}
\usepackage{booktabs} 

\makeatletter
\newcommand\footnoteref[1]{\protected@xdef\@thefnmark{\ref{#1}}\@footnotemark}
\makeatother
\usepackage[para]{footmisc}

\begin{document}

   \title{3D structure of the Milky Way out to 10 kpc from the Sun}
  \subtitle{Catalogue of large molecular clouds in the Galactic Plane}
   \author{Sara Rezaei Kh.
          \inst{1,2}
          \and
          Henrik Beuther
          \inst{1}          
          \and
          Robert A. Benjamin
          \inst{3}
          \and
          Anna-Christina Eilers
          \inst{4}
          \and
          Thomas Henning
          \inst{1}
          \and
          Maria J. Jim{\'e}nez-Donaire
          \inst{6,7}
          \and
          Marc-Antoine Miville-Deschênes
          \inst{5}
          }

   \institute{Max-Planck-Institute for Astronomy, Königstuhl 17, 69117 Heidelberg, Germany\\
              \email{s.rezaei.kh@gmail.com}\\
         \and
             Chalmers University of Technology, Department of Space, Earth and Environment, SE-412 93 Gothenburg, Sweden
        \and
            Department of Physics, University of Wisconsin-Whitewater, Whitewater, WI 53190, USA
        \and
            Physics Department and Kavli Institute for Astrophysics and Space Research, Massachusetts Institute of Technology, 77 Massachusetts Ave, Cambridge MA 02139, USA 
        \and
            Institut d'Astrophysique Spatiale, CNRS, Univ. Paris-Sud, Université Paris-Saclay, Bât. 121, F-91405 Orsay, France
        \and
           Observatorio Astronómico Nacional (IGN), C/Alfonso XII 3, 28014 Madrid, Spain 
        \and
            Centro de Desarrollos Tecnológicos, Observatorio de Yebes (IGN), 19141 Yebes, Guadalajara, Spain
             }
   \date{Received -----; accepted -----}

  \abstract
  {Understanding the 3D structure of the Milky Way is a crucial step in deriving properties of the star-forming regions, as well as the Galaxy as a whole. We present a novel 3D map of the Milky Way plane that extends to 10 kpc distance from the Sun. We leverage the wealth of information in the near-IR APOGEE dataset and combine that with our state-of-the-art 3D mapping technique using Bayesian statistics and the Gaussian process to provide a large-scale 3D map of the dust in the Milky Way. Our map stretches across 10 kpc along both the X and Y axes, and 750 pc in the Z direction, perpendicular to the Galactic plane. Our results reveal multi-scale over-densities as well as large cavities in the Galactic plane and shed new light on the Galactic structure and spiral arms. We also provide a catalogue of large molecular clouds identified by our map with accurate distance and volume density estimates. Utilising volume densities derived from this map, we explore mass distribution across various Galactocentric radii. A general decline towards the outer Galaxy is observed, followed by local peaks, some aligning with established features like the Molecular Ring and segments of the spiral arms.
  Moreover, this work explores extragalactic observational effects on derived properties of molecular clouds by demonstrating the potential biases arising from column density measurements in inferring properties of these regions, and opens exciting avenues for further exploration and analysis, offering a deeper perspective on the complex processes that shape our galaxy and beyond.}

    \maketitle
%

\section{Introduction}
The multi-physics, multi-scale nature of star formation is the centre of today’s star formation challenges. Understanding the gathering of material from small scales in protoplanetary discs to molecular clouds, and accumulation in galaxies are key to providing a holistic picture of star formation \citep{Kennicutt12,Padoan14,Krumholz18}.

Star-forming regions contribute to the overall evolution of galaxies and different galactic environments affect the formation and evolution of star-forming regions. The structures of the galactic disc components, such as spiral arms, influence the distribution and evolution of gas and dust, and therefore star formation in a galaxy. 
Spiral galaxies exhibit active star formation within their spiral arms composed of a concentration of gas and dust. To understand the role of spiral arms in star and galaxy formation and evolution, knowledge of the location of the arms, as well as their components is of utmost value. Resolving individual stars formed within their birth environment, as well as their position within the large-scale galactic environments turns the Milky Way into a unique laboratory with the current observational techniques. However, our position within the dusty disc of the Milky Way has long limited our understanding of the location and substructures of star-forming arms of the Milky Way to the 2D plane-of-the-sky views and uncertain kinematic distances.

The spiral nature of the Milky Way was initially identified in the 1950s through the determination of distances to objects emitting emission lines and the discovery of 21 cm radio observations \citep{Oort52,Hulst54,Morgan55}. Since then, numerous works have focused on characterising the positions of spiral arms in the Milky Way via various approaches; from studying young stars \citep{Russeil03,Cantat18,Romero19}, to atomic and molecular gas kinematics \citep[][]{Drimmel01,Kalberla09,Dame01,Roman10,Miville17}, and maser parallax measurements \citep{Reid19}. However, despite all improvements, due to challenges in estimating distances and obscuration caused by line-of-sight (LOS) extinction, an accurate picture of the exact structure of our galaxy remains elusive to this date.

Multi-scale 3D maps of the Milky Way are not only important from the Galactic perspective, but they also provide the primary steps required to connect the Galactic and extragalactic star formation studies. Owing to the new developments in extragalactic observations, recent studies of external galaxies have reached the resolutions of individual large molecular clouds \citep[tens of parsecs; e.g.][]{Schinnerer13,Leroy21,Sun20,Leroy21}. \cite{Kainulainen22} showed the substantial effects that the viewing angle can have on the estimated properties of the molecular clouds. A face-on view of the Milky Way taken from the 3D maps further allows for studying the effects of observations on derived star formation properties of external galaxies, such as the observed aperture size, inclinations and viewing angles, and scale height on the obtained physical properties.

Since its launch in 2013, the European Space Agency’s Gaia mission has revolutionised Milky Way studies by providing accurate astrometric measurements to individual sources \citep{Gaia_collaboration16}. The latest Gaia data release (Gaia DR3) in 2022 contained full astrometric solutions for nearly 1.5 billion sources with a magnitude limit of G=21 \citep[Gaia Collaboration,][]{Brown21}. The parallax estimates of Gaia with micro arcsecond precisions are the main source of recent 3D developments in the Milky Way studies. The Gaia satellite therefore offers an ideal set of data for studying nearby individual molecular cloud substructures in 3D \cite[e.g.][]{Grossscheld18,Rezaei_Kh_18a,Rezaei_Kh_20,Rezaei_Kh22,Zucker21}, as well as the 3D structure of the local Milky Way \cite[e.g.][]{Green_19,Leike_20,Vergely_22,Eden_23}. Furthermore, the precise 3D positions of a wide range of stellar types observed by Gaia, allow the association of stars with different masses and ages to various cloud components in the ISM to study the evolutionary stages of molecular clouds \citep{Rezaei_Kh_20}. However, due to the optical nature of the Gaia observations, the studies remain limited to nearby ($<$ $\sim$3 kpc) regions. Therefore, to study the large-scale physics of the ISM, complementary near-infrared (IR) datasets are of great importance. In a pilot study in \cite{Rezaei_Kh_18b}, we showed the strength of the near-IR data as great tools for approaching far distances in the Galactic plane. In this work, we showcase the power of near-IR data combined with machine learning techniques to provide a novel 3D map of our Galaxy that expands out to 10 kpc.

The paper is organised as follows: we briefly summarise our 3D mapping technique and the dataset used in this work in section \ref{sec:method}. We then present the 3D map of the Milky Way and explain its features in section \ref{sec:results}, followed by the catalogue of large molecular clouds from our map. In section \ref{sec:discussion}, we compare our map to existing CO and maser observations and discuss the distribution of the clouds in the Galaxy. Additionally, we discuss our findings in the context of extragalactic studies in section \ref{sec:galaxies}. 

The catalogue of selected large molecular clouds (table \ref{tab:clouds}) and the full 3D map can be accessed online with the paper. The users are advised to read the caveat section before using the full 3D map.

\section{Data and methods}
\label{sec:method}
In this section, we explain the input data and the technique used for the production of our 3D map.
\subsection{3D mapping technique}
\label{sec:m_technique}
Our 3D mapping technique has been extensively explained in \cite{Rezaei_Kh_17,Rezaei_Kh_18b}, with further changes and improvements explained in \cite{Rezaei_Kh_20}. Here we briefly summarise the main aspects of our 3D analysis.
Our technique uses the 3D positions of the stars (l,b,d) and their LOS extinction as the input data. It then divides the LOS of each star into small 1D cells in order to approximate the observed extinction toward each star as the sum of the dust in each cell along its LOS. After having done that for all observed stars, our likelihood is formed. The model then takes into account the neighbouring correlation between all points in 3D using the Gaussian Process; i.e. the closer two points in the 3D space, the more correlated they are. This is our prior. Having prepared both the Likelihood and the Prior, the model uses extensive linear algebraic analysis to determine the probability distribution of dust density at any arbitrary point in the observed space, even along the LOS that was not originally observed.\\
Our 3D mapping technique consists of the following unique features:
\begin{itemize}
    \item[$\bullet$] It accounts for both distance and extinction uncertainties in the input data. As a result, our input is not limited to strict data quality cuts and can leverage more observed stars.
    \item[$\bullet$] Owing to the Gaussian-Process based 3D spatial correlation, the final results are smooth and clear of LOS elongated artefacts, also known as ``fingers-of-god'' effects.
    \item[$\bullet$] The predicted dust densities are analytically calculated, therefore, our results are devoid of biases and artefacts often caused by incorrect use of Gaussian Process approximations. Our predictions, thus, can be traced back to the input data.
    \item[$\bullet$] Both the mean and standard deviation of the predicted densities are calculated analytically, thus the precision and reliability of our predictions can be directly evaluated, which increases the robustness of our analysis.
\end{itemize}
The model incorporates several hyper-parameters determined by the input data \citep[see][for more details]{Rezaei_Kh_17,Rezaei_Kh_18b}. One of these parameters is the cell size, which is tuned based on the typical spacing between input stars and serves as the minimum resolution for the final map. In areas with denser and more informative data, the map's resolution will be higher. Another parameter is the correlation length, which defines the range over which spatial correlations exist. Typically, it is a few times the cell size to ensure the connection between nearby cells in 3D. The third hyperparameter, known as the scale variance, is calculated using the data's amplitude and uncertainties, reflecting the variance of predictions. After introducing the datasets in the following section, we will detail these parameters for our models.
\subsection{Input data: Distance and Extinction estimates}
\label{sec:m_data}
%
\begin{figure}
    \centering
    \includegraphics[width=0.49\textwidth]{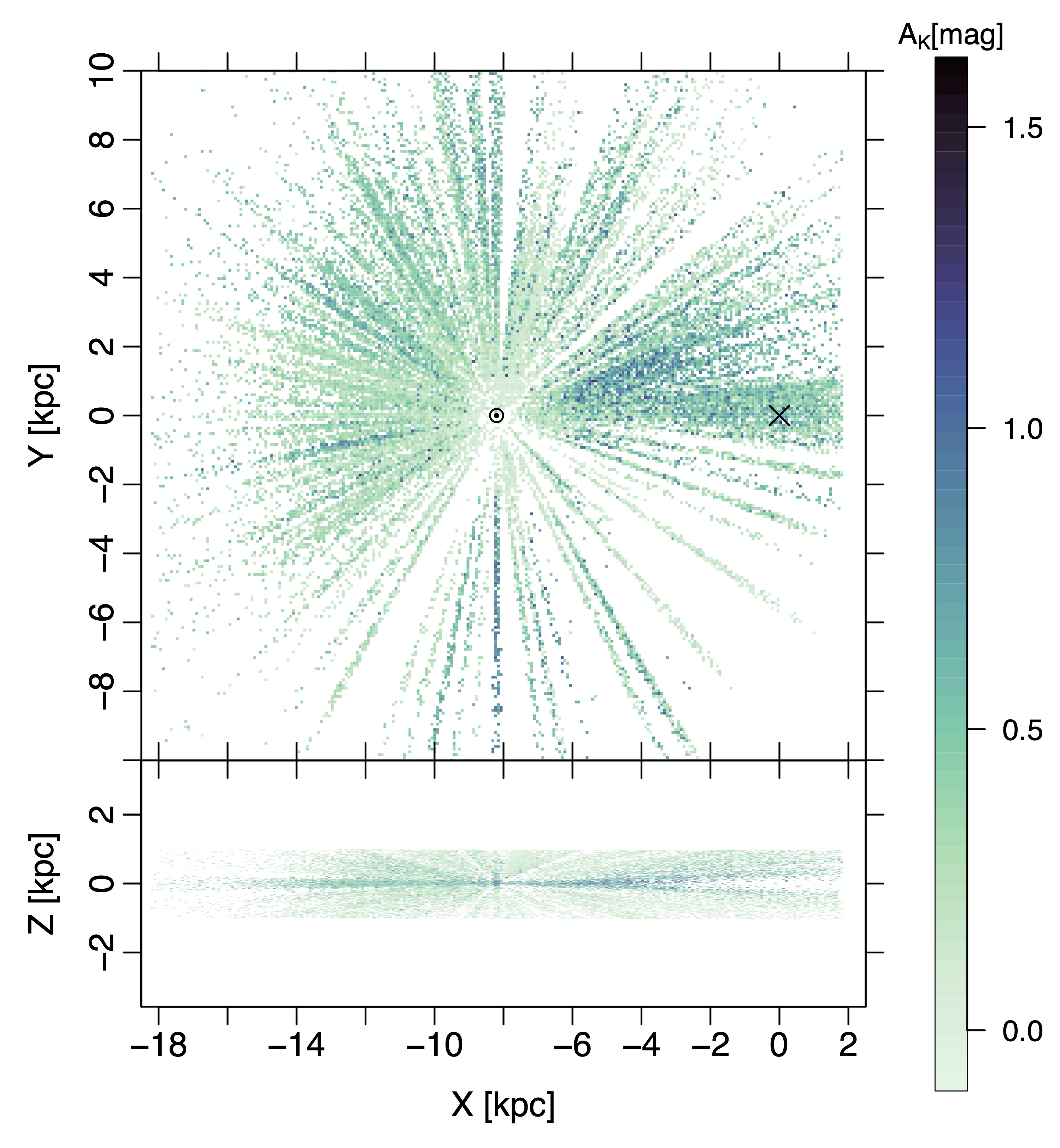}
    \caption{Final sample used as our input data. The top panel shows the X-Y plane, perpendicular to the Galactic plane and the bottom panel presents the X-Z plane where our cut on the 1kpc Galactic height is visible. The colour shows estimated extinctions for individual stars. The Sun is at (-8.2,0,0) and the Galactic Centre is marked with an X, assuming the Sun is at the distance of 8.2 kpc from the Galactic Centre. The gaps between different LOS indicate that the sky is not observed uniformly by APOGEE.}
    \label{fig:input}
\end{figure}
The Sloan Digital Sky Survey IV’s Apache Point Observatory Galactic Evolution Experiment \citep[APOGEE-2,][]{Blanton17,Majewski17,Abolfathi17} is a near-IR high-resolution spectroscopic survey targeting bright stars \citep{Eisenstein11,Zasowski13}. In the near IR, the effects of extinction are about an order of magnitude lower than at optical wavelengths, enabling APOGEE to observe stars in the highly obscured regions of the Galactic disc and towards the Galactic centre. APOGEE does not survey the sky uniformly but rather targets cool stars, particularly red giants, through multiple components of the Galaxy including thin and thick discs \citep{Eisenstein11,Zasowski13}. The 16th data release of APOGEE published in 2020 \citep{APOGEE16}, covers different parts of the Galactic plane out to distances beyond the Galactic centre and includes sources from the southern hemisphere \citep{APOGEE16}. This allows us to map the Milky Way plane with resolutions down to giant molecular cloud sizes ($\sim$ 100 pc) and probe the large-scale structure of the Galaxy, such as spiral arms, and the properties of the large molecular clouds to unprecedented accuracies to date.

Extinction measurements come directly from the APOGEE pipeline: all APOGEE sources have corresponding observations in multi-band photometry in the near- and mid-IR with the Two Micron All-Sky Survey \citep[2MASS,][]{Skrutskie06} and the Wide-Field Infrared Survey Explorer \citep[WISE,][]{Wright10} respectively. Therefore their extinctions are easily estimated using the Rayleigh-Jeans Colour Excess Method \citep[RJCE,][]{Majewski11}.\\
As mentioned in section \ref{sec:m_technique}, in order to infer the 3D distribution of the dust, our method requires the 3D positions of stars. The distance estimates for the APOGEE sources are calculated using spectrophotometric parallaxes computed based on the method of \cite{Hogg19}. The approach leverages a data-driven model that combines photometric and spectroscopic data, aiming to describe the parallaxes of giant stars. It employs a feature vector containing photometric and spectroscopic information, resulting in a 7460-dimensional feature space. The optimization process considers the uncertainties in Gaia parallax measurements and an offset is applied to account for known parallax biases. Given the sparsity of information within APOGEE spectral pixels, a regularization term is introduced to enhance the model's accuracy. The study achieves a median relative uncertainty in spectrophotometric parallax of $\sim8\%$, a significant improvement compared to Gaia parallax, especially for stars beyond a heliocentric distance of 3 kpc \citep{Hogg19,Ou_23}. For a more comprehensive understanding, readers are encouraged to refer to the original paper. This enhanced accuracy enables the mapping of the Milky Way up to distances beyond the Galactic centre.

Having obtained distance and extinction estimates, we have all the essential input data needed for our model. Given our specific interest in the structure of the Galactic Plane, and considering that the majority of APOGEE observations are designed for these regions, we narrow down our input data to include only absolute Galactic heights below 1 kpc, allowing us to focus on the Milky Way midplane. Figure \ref{fig:input} shows the input data used in our model. While the distance estimates for APOGEE sources extend beyond 10 kpc from the sun, the density distribution of stars, as illustrated in Fig.\ref{fig:input}, notably decreases as we approach the 10 kpc distance. As a result, we confine our map to the 10 kpc range along the X- and Y- axes. The final sample contains more than 44\,000 stars and the maximum K-bank extinction in the sample is $\sim$ 1.6 magnitudes. We also note that the number of stars drops significantly in the inner $\sim$500 pc, therefore we limit our predictions to distances beyond 1 kpc.

As mentioned in section \ref{sec:m_technique}, the hyper-parameters of the model are set according to the input data. For the APOGEE data used in this work, the parameters are as follows: cell size of 200 pc, correlation length of 1 kpc, and scale variance of 5e-09 $pc^{-2}$. Additionally, for all plots and analysis in this work, we assume the distance of the Sun from the Galactic Centre of 8.2 kpc and the height from the Galactic midplane of 7 pc \citep[in agreement with][]{Reid19,Gravity_21,Darling_23,Leung_23}.

\section{Results}
\label{sec:results}
Using the input data (as explained in section \ref{sec:method}), we produce the 3D map of the dust in the Galactic plane for a radius of 10 kpc in X-Y plane and 750 pc in the Z direction, perpendicular to the Galactic plane. Our predictions are made for points on a regular grid of 100 pc in X, Y, and for 5 layers in the Galactic heights of -750, -375, 0, 375, 750 pc. Given the sparsity of the APOGEE data, which sets the model's hyper-parameters, especially in regions away from the Galactic midplane, predicting on a denser grid in either X, Y, and Z directions does not add further useful information to the map.
\subsection{Features of the map}
\label{sec:features}

\begin{figure*}[h!]
    \centering
    \includegraphics[width=\textwidth]{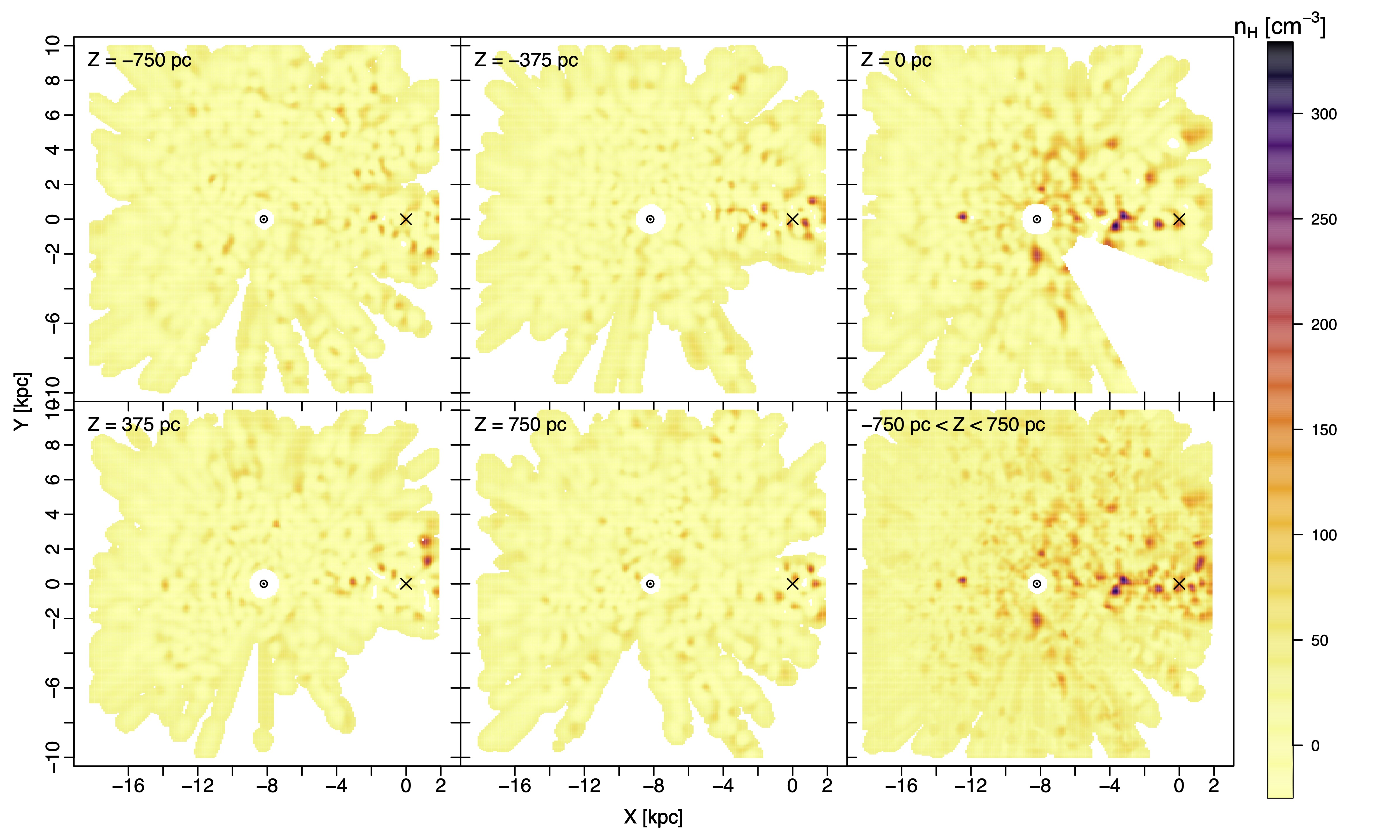}
    \caption{Face-on view of our 3D map of the dust in the Galactic plane for different Galactic heights. The bottom right panel shows the combined map from -750 pc to 750 pc in the Galactic height. The colour shows the mean of our predicted density (cm$^{-3}$) for each pixel (100$\times$100 in X-Y plane) in all panels except for the bottom right where the colour represents the maximum density within the 1.5 kpc height. The Sun is at (-8.2,0,0) and the Galactic Centre is marked with a $\times$, assuming the Sun is at the distance of 8.2 kpc from the Galactic Centre. White regions are areas devoid of input data.}
    \label{fig:3D_mosaic}
\end{figure*}

Figure \ref{fig:3D_mosaic} shows our 3D map for different Galactic heights, as well as a combined map. The white areas are regions devoid of input stars; this is particularly visible in parts of the fourth quadrant of the Galactic midplane (Z=0) where, as seen in Fig. \ref{fig:input}, our current input lacks data. The densities are converted from our model's units of $mag/pc$ to cm$^{-3}$ assuming $A_K/N_H = 0.7 \times 10^{-22}$  $cm^{2} mag/H$ \cite{Draine_09}. Numerous dust density substructures are visible in Fig. \ref{fig:3D_mosaic} in various distances and heights; the majority of which appear, as expected, in the Galactic midplane (z=0). In particular, there are multiple high-density clouds towards the Galactic Centre, with the densest structures of the map appearing at a Galactocentric radius of about 4 kpc, likely associated with the so called Molecular Ring \citep{Krumholz_05}. We do not see major differences between positive and negative heights in terms of the presence of the warp in the Galaxy. This could be due to the incompleteness in the range covered by the input data, or the presence of the stellar warp in further distances, as suggested by \cite{Poggio_20}.

\begin{figure*}[h!]
    \centering
    \includegraphics[width=\textwidth]{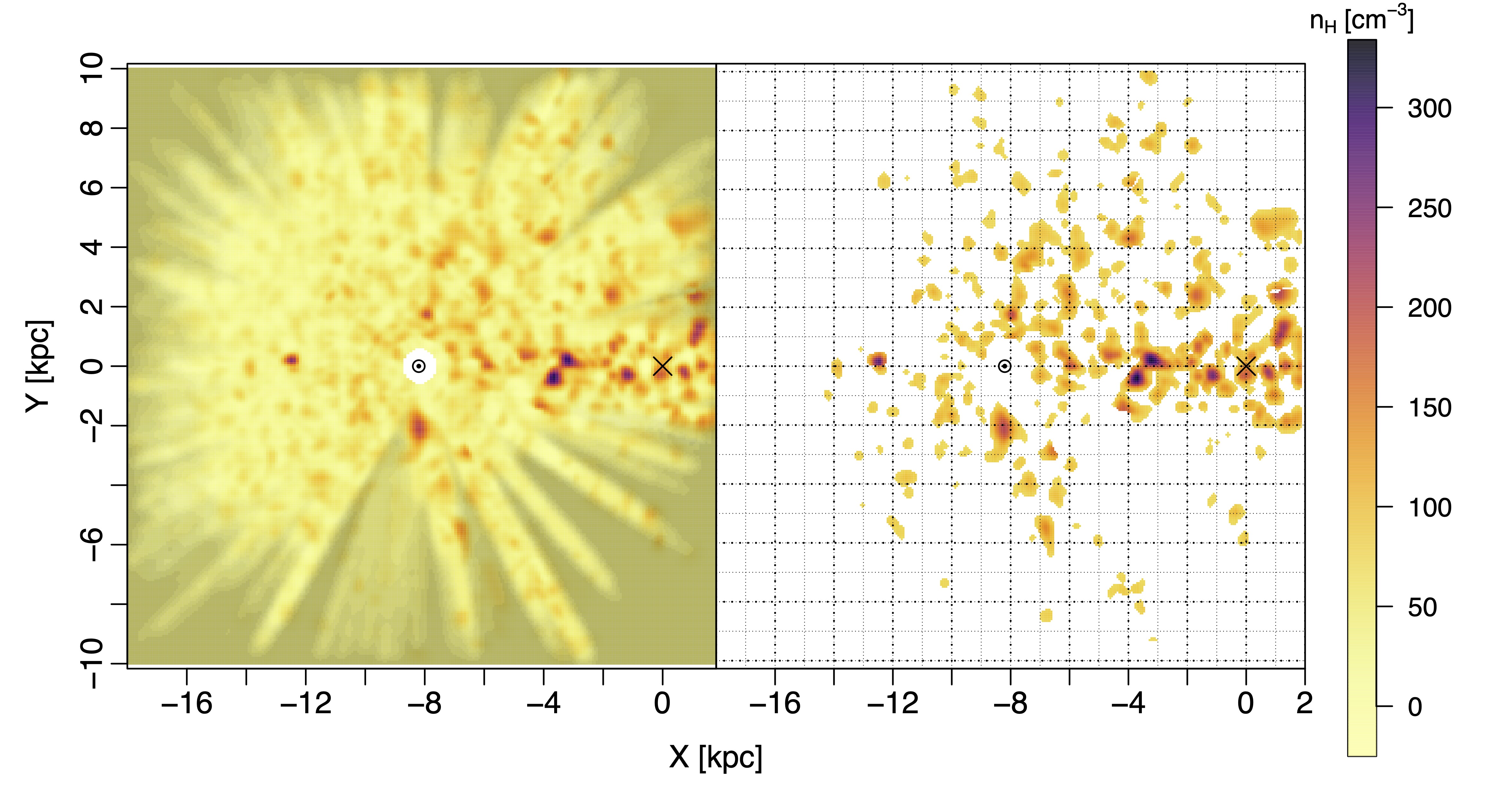}
    \caption{Face-on view of the 3D map for all Galactic heights. Left panel: same as the bottom right panel of Fig. \ref{fig:3D_mosaic}, in addition to having shaded areas illustrating regions of high uncertainty (fractional uncertainties larger than 50\%). Right panel: clouds selected from the 3D map whose density values lie three standard deviations above the mean of the Gaussian Process. This corresponds to densities above 80 cm$^{-3}$.}
    \label{fig:3D_all}
\end{figure*}

As explained extensively in our previous works \cite{Rezaei_Kh_17,Rezaei_Kh_18b,Rezaei_Kh_20} and in section \ref{sec:m_technique}, the robustness of the 3D dust maps, especially those based on Gaussian Processes, lies within the reliable calculation of the uncertainties. Therefore, we use our predicted mean densities together with their uncertainties, which are calculated analytically. Figure \ref{fig:3D_all}, left panel, shows our 3D map with shaded areas representing regions of large uncertainties (fractional uncertainties larger than 50\%). A separate uncertainty map is also shown in Appendix A. The smallest fractional uncertainty is about 5\% and belongs to high-density regions. The radial patterns in the shaded plot clearly show areas with a lack of input data, as seen in Fig. \ref{fig:input}. 

The right panel of Fig. \ref{fig:3D_all} shows clouds extracted from the 3D map with statistically significant densities. Statistically significant in this context means three standard deviations above the mean of the Gaussian process that is used for density calculations. It is important to note that this does not imply three sigma above the ``noise'', which in this case would be around zero, but refers to a much higher threshold: the Gaussian process uses a mean density for each region based on the input stellar extinction, which is already higher than the average ``noise''. In order to make sure the clouds are ``real'' and their densities are not derived by the mean density of the Gaussian Process, we go three standard deviations above this value to have a pure sample of dense clouds. This corresponds to densities above $\sim$80 cm$^{-3}$. The fractional uncertainties of the selected clouds are between 5 and 30 percent. The first evaluation of the map and the clouds within it does not show a clear indication of the spiral arms in the Milky Way. The local arm and segments of the Perseus arm are the only clear arm features that could be extracted from the map. We explore these further in the discussion section (section\ref{sec:discussion}).

From our map of clouds (Fig. \ref{fig:3D_all}, right), we select regions with densities above 100 cm$^{-3}$, corresponding to the molecular phase, and provide a catalogue of large molecular clouds in the Milky Way (table \ref{tab:clouds}). We deliver accurate distance estimates to the centre of each cloud, the uncertainty of the estimated distance, the extent of the cloud, its mean density and standard deviations, and its association with known star-forming regions and spiral arms. 

\begin{figure}
    \centering
    \includegraphics[width=0.49\textwidth]{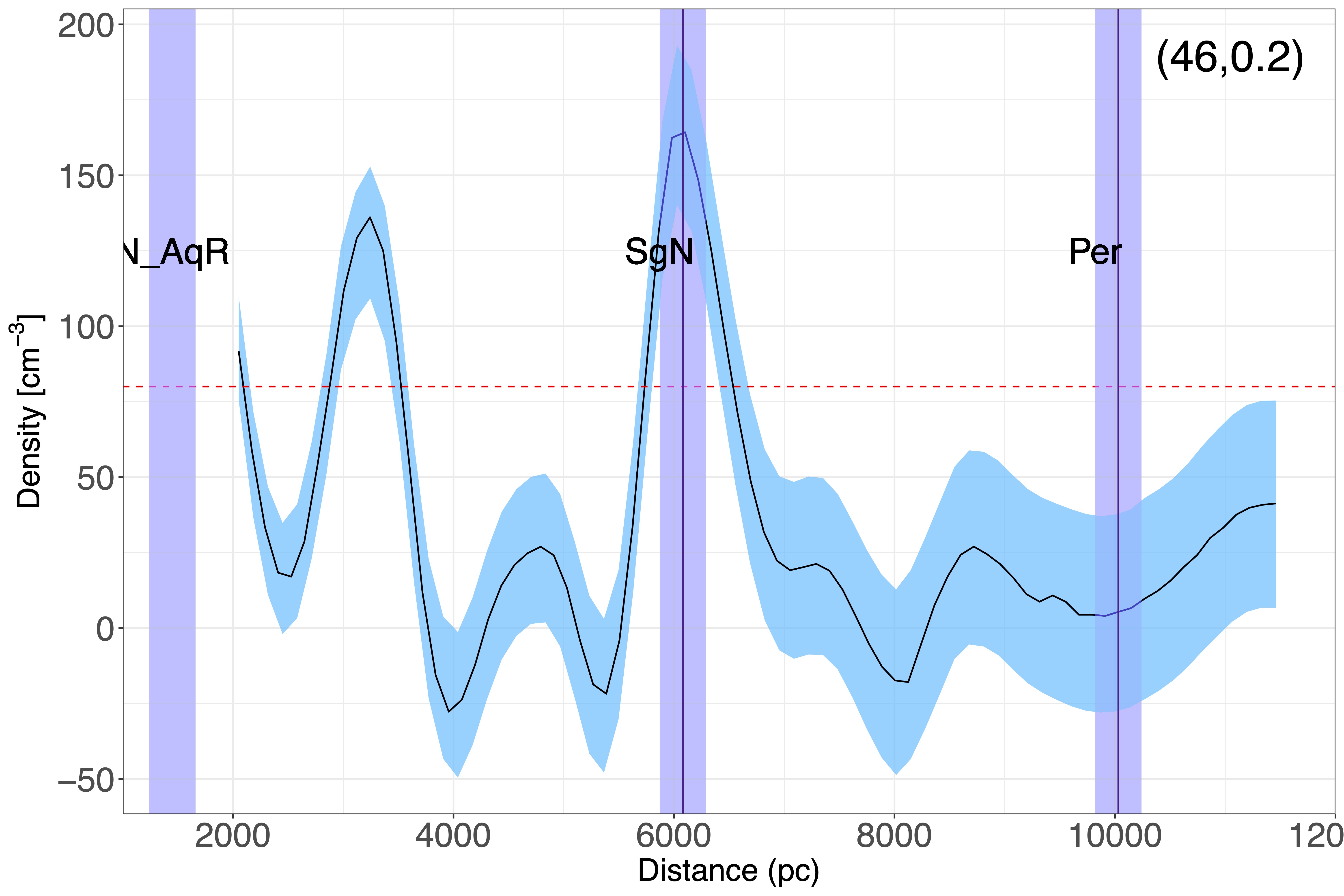}
    \caption{An example of how the clouds in Table \ref{tab:clouds} are assigned to known structures and spiral arms. The black line with the blue shaded uncertainties are our density predictions as a function of distance for a given LOS and the purple vertical lines are all probable distance estimates from the BESSEL survey \citep{BESSEL_16} for the same LOS with velocities observed in \cite{Dame_86} (here are Aquila rift, Sagittarius Near, and Perseus, for example). The dashed red line indicates our threshold for selecting statistically significant clouds. If there is a match, like the Sagittarius Near arm in the middle, we assign the cloud to that particular structure. The LOS' galactic longitude and latitude are shown in the top right corner. }
    \label{fig:los}
\end{figure}

The associations with spiral arms and star-forming regions have been determined using a combination of catalogues and studies: 
in the first quadrant, we used the catalogue of molecular clouds from \cite{Dame_86} which has a similar resolution to our map, combined with the BESSEL distance calculator \citep{BESSEL_16}. The BESSEL survey uses a combination of maser parallaxes, spiral arm models based on masers, and kinematic distances to give all probable distance estimates for a given LOS and velocity (\textit{l,b,v}). For each cloud from \cite{Dame_86} with a matching LOS to a cloud centre from our map, we use the BESSEL distance calculator to get all probable distance estimates for that (\textit{l,b,v}). For a given LOS, if we had a peak in density in our 3D map (within 0.5 degrees from that LOS) that matches one of the probable distance estimates from the BESSEL survey for that (\textit{l,b,v}), we assume they correspond to one another (see Fig. \ref{fig:los} for an example). For the rest of the map, we use the catalogue of radio sources from \cite{Westerhout_58} associated with star-forming regions, the catalogue of H{\ensuremath{\alpha}}-emission regions in the southern Milky Way by \cite{Rodgers_60}, and the catalogue of star-forming regions by \cite{Binder_18}. For the arm associations, we used masers from \cite{Reid19}.

In addition to the over-densities and spiral arm segments, one clear feature of the map is the presence of large cavities. These cavities are marked by dashed lines in Fig. \ref{fig:3D_compare}. While we were able to extract a clear sample of clouds with reliable densities from the map, differentiating real cavities from regions of underestimated densities due to missing data is very difficult. As a result, we limit our sample of the cavities to nearby regions (within $\sim$ 4 kpc from the Sun) to avoid mistakenly categorizing regions without complete input data as cavities. This would also allow comparison to other 3D maps with overlapping regions. 

\subsection{Comparison to other 3D maps}
\label{sec:3d_comparison}

\begin{figure*}[h!]
    \centering
    \includegraphics[width=\textwidth]{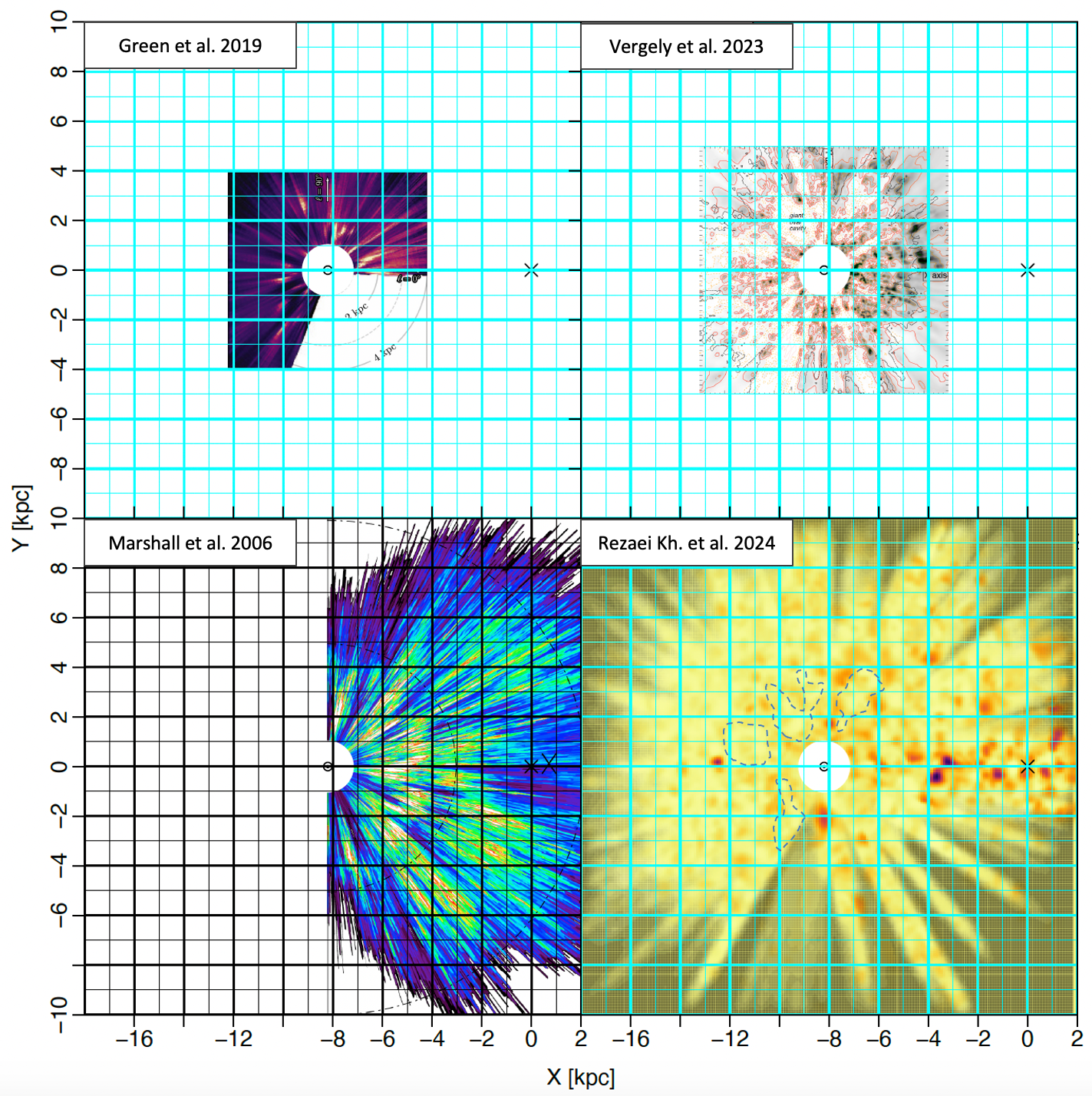}
    \caption{Qualitative comparison between 3D dust maps of the Galactic plane. Grids are drawn for easier comparisons between maps. The lower right panel shows our 3D map where dashed lines represent identified large cavities. Similar cavities are seen in the upper right panel representing the work of \cite{Vergely_22}. While similar cavities seem to be present at similar locations in the other two panels, \citep{Green_19,Marshall06}, due to substantial artefacts, a direct comparison appears difficult. Apart from our map, it is only the \cite{Marshall06} work that expands beyond the Galactic Centre; despite the elongated artefacts, multiple over-densities in the first quadrant and towards the Galactic Centre are evident in both maps.}
    \label{fig:3D_compare}
\end{figure*}

\cite{Vergely_22} has provided one of the most promising maps of the Galaxy by combining Gaia parallaxes with the cross-match of the Gaia data with 2MASS and WISE photometry. They also consider a full 3D correlation in space to provide an artefact-free, smooth 3D map of the Milky Way out to 4.5 kpc from the Sun at a resolution of 25 pc \citep{Vergely_22}. \cite{Green19} use multiband photometry from PANSTARRS together with the Gaia parallaxes to simultaneously derive distance and extinction to individual stars. Their map has a high resolution on the plane of the sky; however, due to the separate treatment of each LOS, elongated artefacts are visible in their final map. The technique of \cite{Marshall06} is particularly different from the others: they measure the colour excess of stars by assuming a Galaxy model and the intrinsic colours of stars. They then use this colour excess to determine a star's distance and extinction. While there exist substantial LOS elongations in the map of \cite{Marshall06}, it is the only one that reaches similar distances to that of our map towards the inner Galaxy. 

Figure \ref{fig:3D_compare} shows a qualitative comparison between our map and 3 other maps each based on different techniques and datasets. The maps also show various resolutions, features, artefacts, and density ranges. As a result, we limit our comparison to dominant features and structures of the maps and avoid quantitative comparisons. Large cavities in our map are marked by dashed lines. While a direct comparison to \cite{Marshall06} and \cite{Green19} maps seem difficult because of the existing artefacts, there are still noteworthy features to consider: the large cavity marked in the first quadrant of our map is clearly visible in the map of \cite{Marshall06}, together with its surrounding over-densities. The cavity in the third quadrant as well as the over-densities adjacent to its left are also visible in \cite{Green19}. Amongst all maps, the one from \cite{Vergely_22} shows the most promising comparison. All four large cavities discovered in our map are clearly visible in \cite{Vergely_22} as well as most of their surrounding over-densities. While our map and that of \cite{Vergely_22} are in good qualitative agreement, there are some differences as well: our map has a much larger distance coverage, and in return, the resolution of \cite{Vergely_22} is on average 4 times better than ours; as a result, they recover much smaller structures than our map can achieve. This is visible, for instance, at $l = 270^{\circ}$ where our map recovers a large over-density around the location of the Vela molecular cloud ((x, y) = (-8, -2) kpc), while around the same region, \cite{Vergely_22} recovers multiple smaller substructures that our map is not able to resolve. 

Another significant aspect of our map involves multiple over-densities observed towards the Galactic centre ($Y\simeq0$). The most prominent dense clouds manifest at Galactocentric radii around 3 and 4 kpc. Additionally, there are several over-densities in this direction at Galactocentric radii of 4.5 and 6 kpc. While \cite{Vergely_22} show numerous smaller clouds in this direction, one particular cloud at Galactocentric radii of about 4.5 kpc appears to align with ours. However, as \cite{Vergely_22} note, the reliability range of their map towards the inner Galaxy is particularly limited to about 4 kpc from the Sun. Conversely, \cite{Marshall06}'s map offers extensive coverage for comparison in the inner Galaxy's direction. Multiple over-densities are visible in \cite{Marshall06} towards the direction of the Galactic centre. Two dense regions around $X = -3$; one on the $Y = 0$ line, and another slightly below, correspond well with our dominant clouds in that region, although at slightly different distances. Similarly, another over-density in their map at $X \simeq -4.5$, just above the $Y = 0$ line corresponds well with ours. Additionally, some clouds identified in the first quadrant of our map align with those in \cite{Marshall06}, although differences in distances are observed due to uncertainties and elongated radial structures in \cite{Marshall06}. While our map remains relatively incomplete in most of the fourth quadrant, a few recovered clouds seem to correlate with those in \cite{Marshall06}'s map as well. 

A notable difference between \cite{Marshall06} and our map lies in the density distribution around the Galactic Centre. While \cite{Marshall06} reveals a cavity around and below the Galactic Centre, our map depicts a few clouds in that vicinity. Understanding the reasons for these differences presents a challenge, but we offer our insights: As seen in Fig. \ref{fig:input}, bottom panel, our input data reveals incompleteness around the Galactic Centre at $b=0$. Consequently, most of the identified over-densities in our map around this area stem from adjacent data at higher or lower latitudes. While we acknowledge data incompleteness around the Galactic Centre, we anticipate that with a more comprehensive dataset covering this region, we would identify more clouds and denser concentrations, rather than the reverse. This is supported by recent studies towards the Galactic Centre and the Central Molecular Zone (CMZ), where an accumulation of molecular gas is observed \citep[for an overview of the CMZ, see][]{Henshaw_23}. The observed under-density in \cite{Marshall06} could similarly result from incomplete data in that region or inaccurate distance estimations for stars due to crowding and confusion. The forthcoming data from SDSS-V \citep{Kollmeier17} holds promise in shedding light on this matter.

\section{Discussion}
\label{sec:discussion}
In this section, we discuss our results in the context of Galactic structure and radio observations, as well as its application in extragalactic studies. 
\subsection{Milky Way structure and masers}
\label{sec:masers}
\begin{figure}
    \centering
    \includegraphics[width=0.49\textwidth]{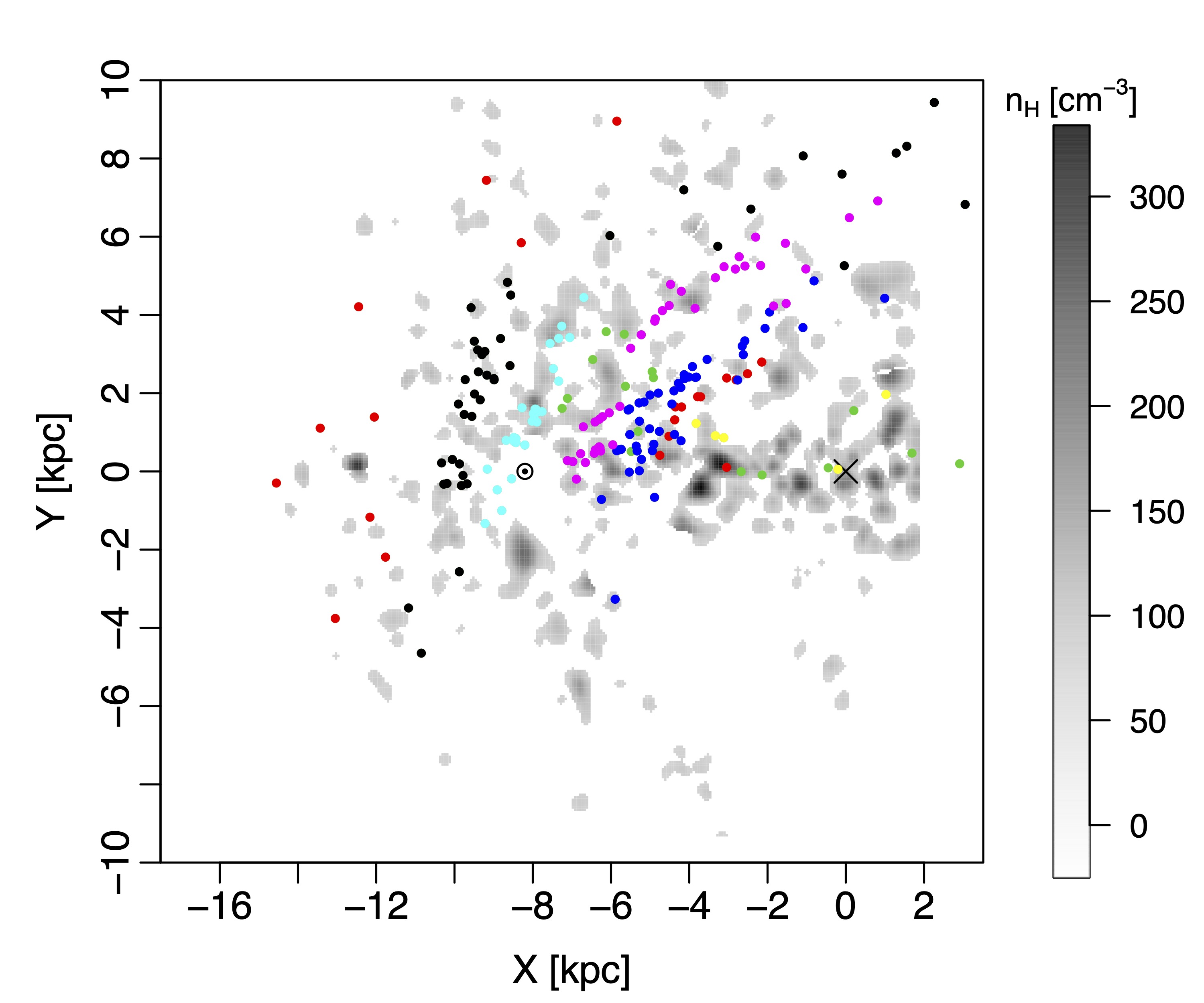}
    \caption{Gray-scale: clouds extracted from our 3D dust map (same as Fig. \ref{fig:3D_all}, right). The colour coding follows the same as Fig. 1 in \cite{Reid19}. Each colour represents masers belonging to a spiral arm; 3-kpc arm: yellow -- Norma–Outer arm: red -- Scutum–Centaurus–OSC arm: blue -- Sagittarius–Carina arm: purple -- Local arm: cyan -- Perseus arm: black. The Green points are equivalent to the white points in \cite{Reid19} illustrating spurs or sources with unclear arm associations.}
    \label{fig:masers}
\end{figure}
We first compare our results with the locations of high-mass star-forming regions identified by trigonometric parallaxes of maser emissions \citep{Reid19}. Mapping the spiral structure of our Milky Way poses significant challenges due to vast distances and dust obscuring the Galactic plane in optical wavelengths. However, using Very Long Baseline Interferometry (VLBI) in radio wavelengths, unaffected by dust, has proven effective in identifying molecular masers linked to young massive stars, providing valuable insights into spiral structures \citep{Reid19}. \cite{Reid19} conducted a comprehensive analysis, gathering around 200 trigonometric parallaxes and proper motions of these masers, primarily from the BeSSeL Survey through VLBA and Japanese VERA project. They associated observed masers with spiral arms by considering patterns in CO and HI Galactic longitude–velocity plots, alongside Galactic latitude information, enabling a better understanding of the Milky Way's spiral structure. 

Figure \ref{fig:masers} shows masers from \cite{Reid19} over-plotted on our clouds extracted from the 3D map (see section \ref{sec:features} and Fig. \ref{fig:3D_all}). There is a significant overlap between our clouds and the masers, particularly at the location of the Local arm, segments of the Perseus arm, as well as Sagittarius–Carina arm. There are several clouds underneath masers belonging to Scutum–Centaurus–OSC, Norma arm, and the spurs in the inner Galaxy; however, it is difficult to establish a one-to-one relation due to the crowding in both masers and cloud distributions. Regardless, the multiple over-densities towards the Galactic centre identified in our map (see section \ref{sec:3d_comparison}), are well represented by the masers in \cite{Reid19}. Additionally, while there seems to be an offset between the masers and our clouds at the location of the Outer arm in the outer Galaxy (red points), our clouds seem to turn and follow the potential spiral arm pattern. This could suggest that the clouds and masers represent different parts of the same spiral arm.

\subsection{Comparison to CO}
\label{sec:co}
\begin{figure*}[h!]
    \centering
    \includegraphics[width=\textwidth]{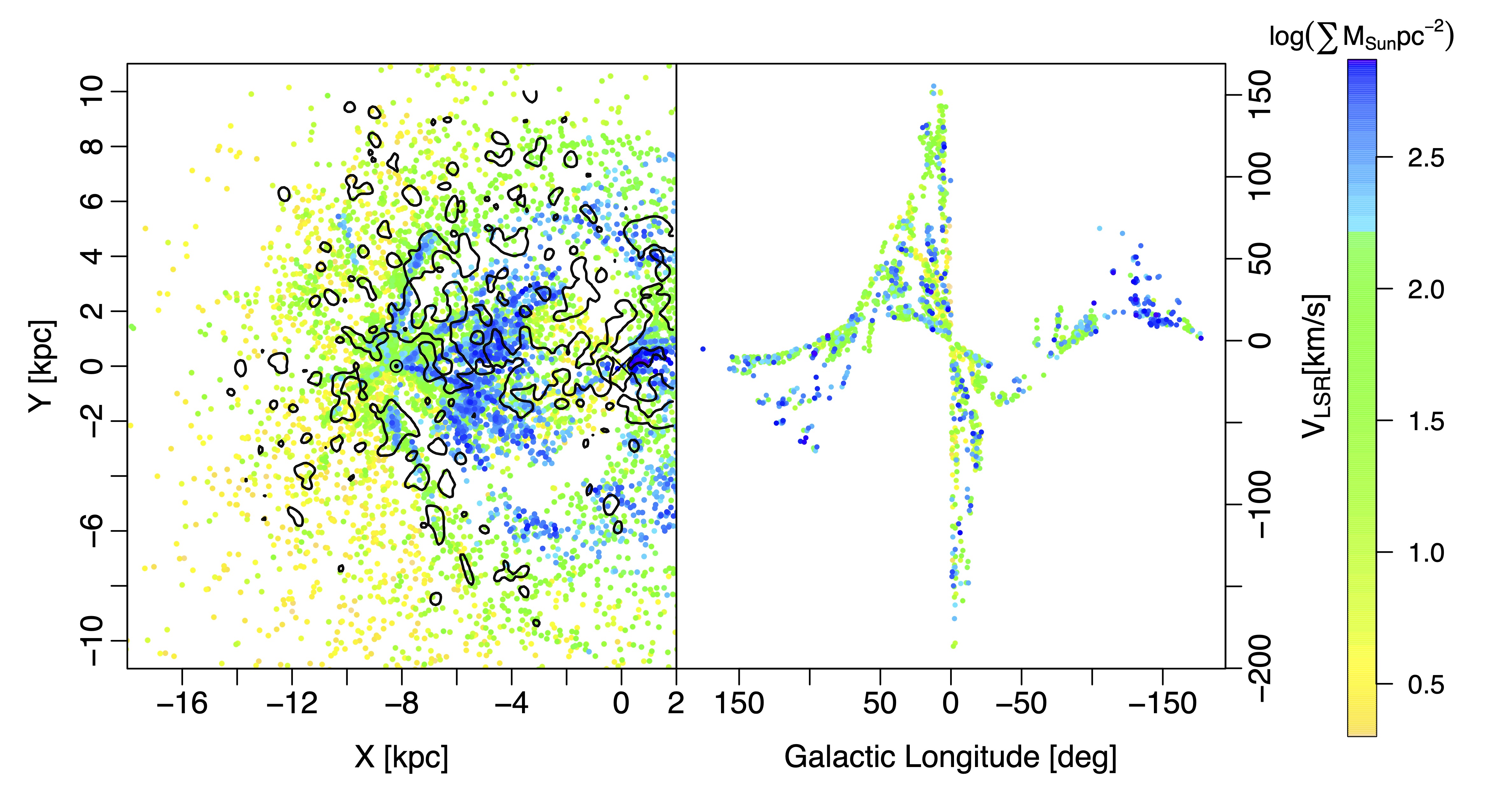}
    \caption{Left panel: 3D distribution of the molecular clouds observed in CO from \cite{Miville17}, with our extracted clouds over-plotted as contours. Right panel: longitude-velocity plot of the molecular clouds from the left, situated underneath our cloud contours. The colour shows the surface density of clouds derived in \cite{Miville17}.}
    \label{fig:CO}
\end{figure*}

\cite{Miville17} provided a catalogue of 8107 molecular clouds spanning the entire Galactic plane and containing 98\% of the observed $^{12}$CO emission within a latitude range of $pm$5 degrees. The catalogue was produced using a hierarchical cluster identification technique applied to the Gaussian decomposition of data initially compiled by \cite{Dame01}. They estimated distances to the clouds using kinematic distance estimates following \cite{Roman-Duval_09} and the rotation curve defined in \cite{Brand93}. They provide physical properties of the clouds including, but not limited to, 3D positions, surface density, physical size, and mass \citep{Miville17}.

We compare our results to their cloud distribution in Fig. \ref{fig:CO}. We first over-plot our clouds as contours on their molecular cloud distribution on the face-on view of the Galaxy. There is a wide circular void in the CO clouds of \cite{Miville17}, particularly noticeable around the Galactic centre extending for about 4 kpc, which is due to limitations in the kinematic distance estimates at these regions. There exists a good agreement between our cloud locations and the CO clouds around the position of the local arm. However, as we get closer to the inner Galaxy and near the Galactic centre, once again it becomes difficult to conclude due to the crowding. It is also important to note that because of the kinematic distance estimates for the CO sources, the distance uncertainties become quite significant in the crowded inner Galaxy, at far distances, in addition to the possible near/far confusion. 

Nevertheless, we proceed with our comparison by extracting CO clouds situated at the same position as our clouds in Fig. \ref{fig:CO}, left, utilising their velocity information to construct a longitude--velocity (L--V) diagram. Fig. \ref{fig:CO}, right, presence patterns on the L-V diagram typically indicative of spiral arms, and commonly used to develop and validate arm models \citep[e.g.][]{Reid19}. The presence of circular motions for the selected molecular clouds is evident from the loop-like trails. Conversely, no clear arm pattern emerges in Fig. \ref{fig:CO}, left, and modelling spiral features from the L--V diagram in Fig. \ref{fig:CO}, right, poses considerable challenges due to unclear and incomplete arrangement of the molecular clouds.

\subsection{Mass distribution in the Galaxy}
\label{sec:Mass}

\begin{figure}
    \centering
    \includegraphics[width=0.49\textwidth]{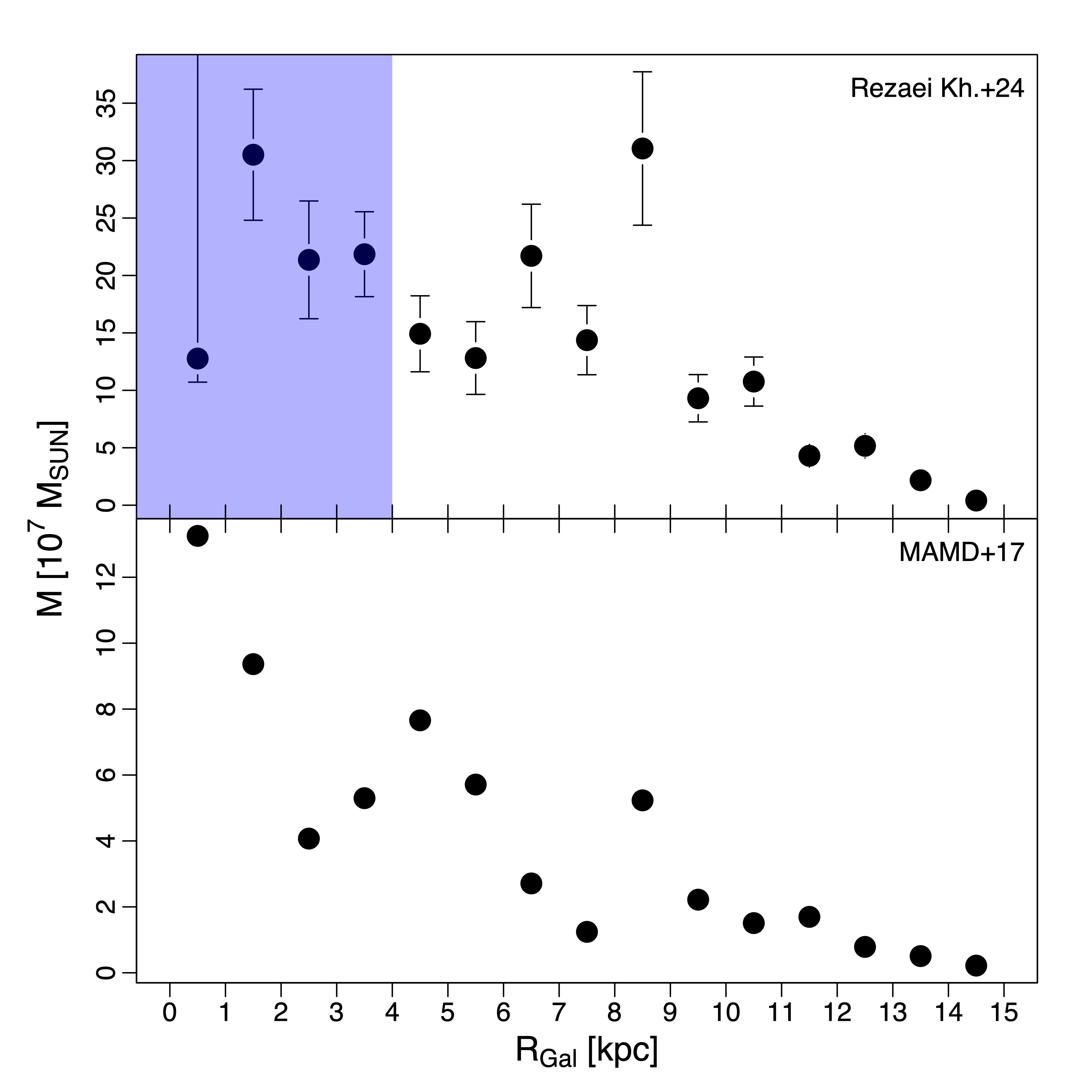}
    \caption{Mass as a function of galactocentric Radius for clouds from our 3D map (top) and CO from \cite{Miville17} (bottom) for galactocentric rings of 1 kpc thickness. In this plot, we have excluded CO clouds from \cite{Miville17} belonging to regions not covered by our map (e.g. parts of the fourth quadrant). The shaded area shows regions where our results are likely underestimated due to the lack of input data. We have not limited the azimuth range for each radial bin; however, due to the limitations in the map, for regions beyond a Galactocentric radius of 2 kpc, our radial average only covers half of a circle (negative X in previous figures).}
    \label{fig:mass}
\end{figure}

Having the 3D distribution of the clouds in the Galactic Plane allows the study of the mass distribution as a function of the galactocentric radius. Fig. \ref{fig:mass} shows the total mass as a function of the galactocentric radius for rings of 1 kpc thickness around the Galactic Centre, derived from our 3D dust volume density map, and that of \cite{Miville17}. To have a fair comparison, we removed clouds from \cite{Miville17} where our map is unable to recover clouds (e.g. parts of the fourth quadrant or at distances far behind the Galactic Centre). For the inner 2 kpc, our map covers the full circle while calculating the mass for the radial bins; however, the outer regions are only averaged over half of a circle (Negative X in the previous plots) because of our limited coverage. It is important to note that due to the lack of data in the Galactic Plane within the inner 4 kpc Galactocentric radii (see Fig. \ref{fig:input}, bottom), we underestimate the total mass in these regions (marked by shaded areas in Fig. \ref{fig:mass}, top).

Overall, the total mass decreases as a function of galactocentric radius, followed by several local peaks. This has already been observed in other studies \cite[e.g.][]{Elia_22,Miville17,Lee_16,Kennicutt12}; however, the location of the peaks in different studies do not always agree. We find the first peak at about 4 kpc, likely associated with the Molecular Ring, predicted by \cite{Krumholz_05}. This peak was reported by \cite{Chiappini_01} at 4 kpc, \cite{Miville17} at 4.5 kpc, and \cite{Elia_22} at 5 kpc. However, as mentioned earlier, our results within the galactocentric radii of 4 kpc should be treated with caution due to incomplete input data in the inner Galaxy. It is also important to note that the dip between 1 and 4 kpc in \cite{Miville17} (Fig. \ref{fig:mass}, bottom), and similar works relying on kinematic distances, is not a real effect but rather due to lack of data in these regions, as illustrated in Fig. \ref{fig:CO} and section \ref{sec:co}. Additionally, we see a local dip between 4-5 kpc followed by a secondary peak at 6.5 kpc, which is in agreement with the 6 kpc peaks of \cite{Miville17,Elia_17}. The 6.5 kpc peak matches the location of the near Sagittarius–Carina arm (see Fig. \ref{fig:masers}). One of the most prominent peaks in Fig. \ref{fig:mass} is the local peak at 8.5 kpc. This matches very well with the local peak of \cite{Miville17} and could be the result of the arrangement of clouds at the local arm and a large segment of the fourth quadrant (see Fig. \ref{fig:CO}, left). There are slight increases at further distances of about 10.5 and 12.5 which could potentially be derived from the concentration of clouds in a segment of the Perseus arm and the Outer arm at these distances; however, the peaks are not as prominent as the previous ones.

\subsection{Impact on extra-galactic studies}
\label{sec:galaxies}
\begin{figure*}
    \centering
    \includegraphics[width=0.49\textwidth]{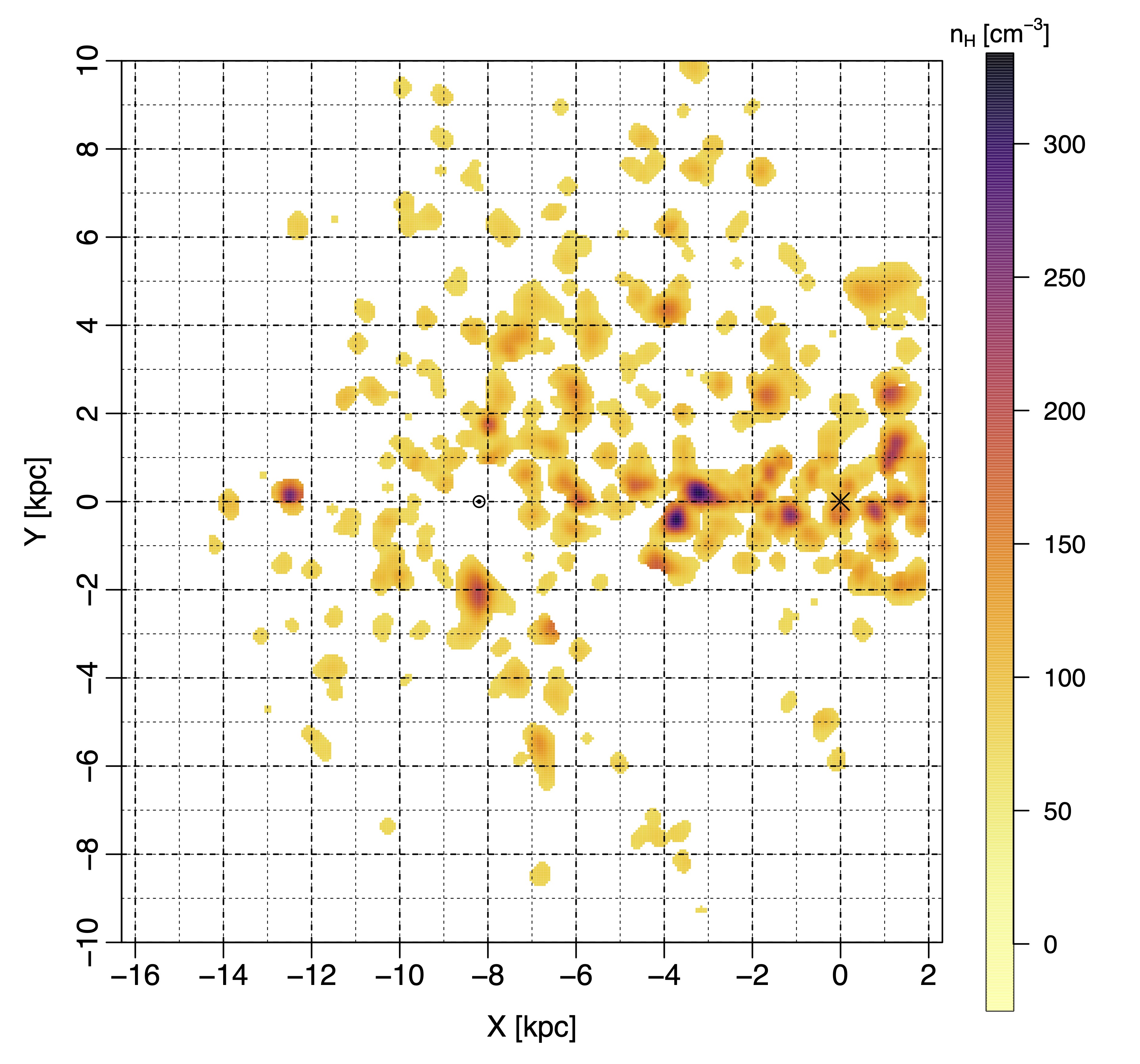}
    \includegraphics[width=0.49\textwidth]{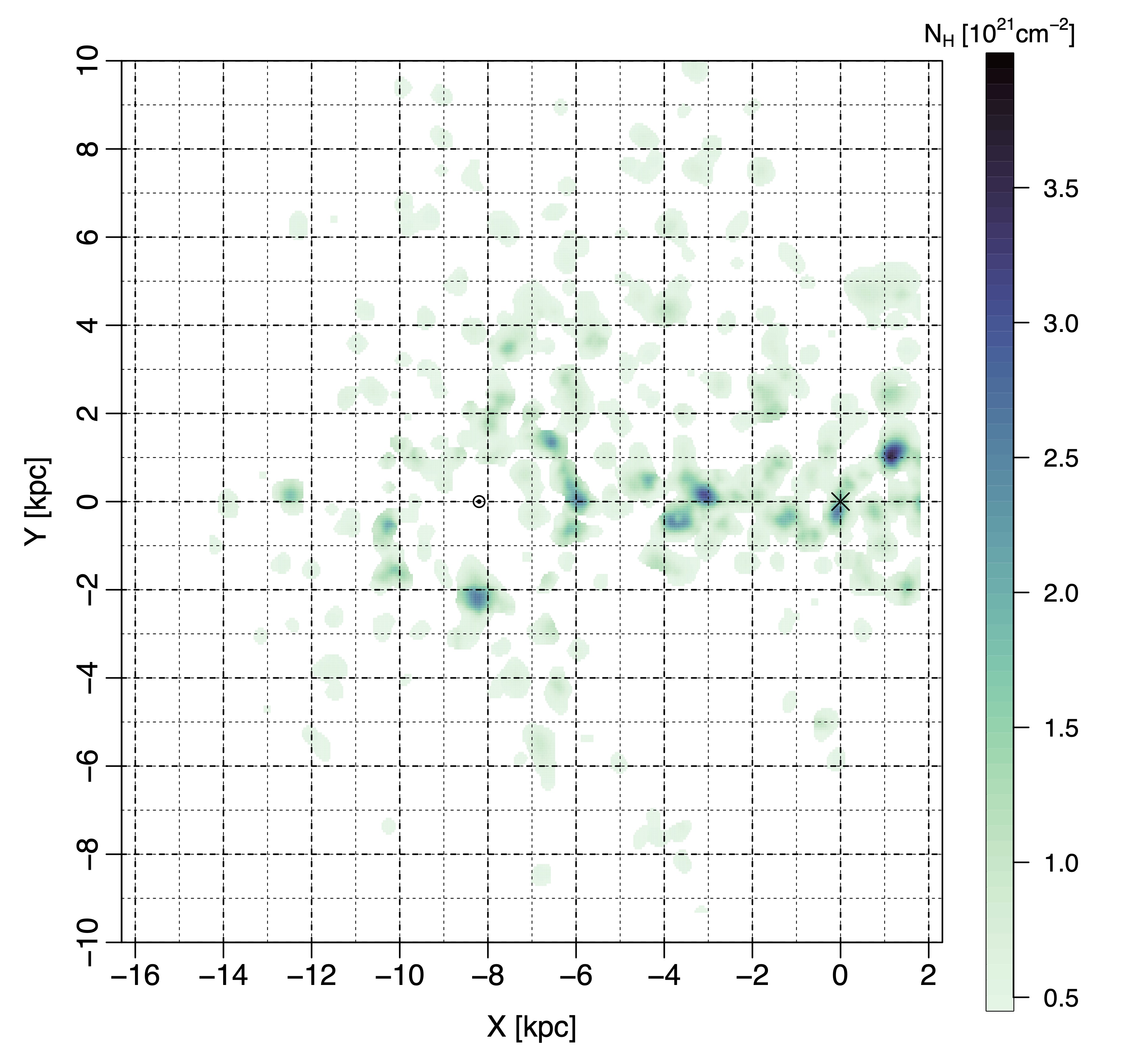}
    \caption{Comparison of the volume density and column density. Left panel: Volume densities of clouds from our 3D map (same as Fig. \ref{fig:3D_all}, right). Right panel: column densities of clouds in left integrated over the Z access.}
    \label{fig:extragalctic}
\end{figure*}
Recent years have seen a surge in studies examining molecular clouds in other galaxies, leveraging our knowledge from the Milky Way to a diversity of extragalactic environments \citep[e.g.,][]{Kawamura_09,Donovan_13,Schinnerer13,Freeman_17,Lee_23}. Large surveys like PHANGS–ALMA \citep{Leroy21} allow us to analyze high sensitivity and resolution observations of giant molecular clouds ($\sim$60-150 pc) across nearby, massive, star-forming galaxies. Understanding the relationship between these clouds and their galactic surroundings is vital to grasp the underlying physics governing their evolution \citep{Sun_22}.

All these observations, however, rely on observed integrated intensities to define and identify discrete molecular clouds and, therefore, can be subject to the projection effects. Observing molecular clouds at large distances increases the probability of having an overlap of different clouds in the same line of sight which, in turn, increases the mean column density of the ensemble substantially. Overlapping clouds in the same line of sight, especially in arm regions, contribute to observed scatter in mass–size relations \citep{Colombo_14,Ballesteros_19}.

Using our 3D map, we can view the Milky Way as an external galaxy and estimate some of the observable parameters of extragalactic studies. Fig. \ref{fig:extragalctic} shows the Milky Way from an external viewpoint with two measured parameters: volume density, not directly accessible to extragalactic Astronomy, and column density, widely used in extragalactic studies to infer properties of the star-forming regions and galaxies. The right panel of Fig. \ref{fig:extragalctic} is derived by integrating our 3D map (left panel) along the Z access (i.e. the Galactic height). Notably, Fig. \ref{fig:extragalctic} indicates that there isn't always a direct correspondence between volume densities (which are directly connected to the physical properties of the clouds) and observed column densities. This occurs due to the buildup of low to moderate densities along the line of sight, creating the illusion of a high-density cloud. This discrepancy holds significant implications for extragalactic studies, particularly when comparing cloud properties across different environments and between galaxies. It not only impacts the location of massive star-forming regions in a galaxy but also affects the total mass derived for each cloud  \citep[see Fig. \ref{fig:extragalctic}, right, and refer to][for more dicussion on the projection effects and mass derivation.]{Rezaei_Kh22,Kainulainen22,Cahlon_24}

The rotational transitions of CO have been the most popular tracers of the bulk molecular ISM in the Milky Way and other galaxies \citep[e.g.,][]{Bally_87,Dame01,Kuno07}. However, CO is quickly thermalized, so its emission does not reflect the different density regimes of a cloud. Constraining the density distribution of star-forming gas in external galaxies is even more challenging since compact, high-density regions within molecular clouds are hard to resolve. High-critical density spectroscopic lines such as HCN(1--0) and HCO$^{+}$(1--0) have become common tracers of dense gas in galaxy disks. Over the last decade, multiple extragalactic surveys at low \citep[e.g. the EMPIRE survey,][]{MJ_19} and high-resolution \citep[e.g.,][]{Gallagher_18,Querejeta_19,Beslic_21} have found systematic trends for the star formation efficiency per unit dense gas as a function of the host galaxy and local environment properties. Nevertheless, these conclusions rest on the ability to translate extragalactic HCN emission into a dense molecular gas mass. In that context, numerous research works have recently shown that these tracers are not as selective of dense gas as previously assumed. Instead, their integrated intensity is dominated by the emission of low-density regions \citep[e.g.,][]{Kauffmann_17,Pety_17,Evans_20,Tafalla_21,Tafalla_23,Dame_23}.

Given the findings of our present study and other Galactic and extragalactic works that concentrate on the LOS superposition effects of multiple clouds and extended objects, affecting the derived properties of the clouds \citep[e.g.][]{Ballesteros_19,Rezaei_Kh22,Kainulainen22}, one way to minimise such biases involves observations of molecular line emissions in critical high-density environments. In the Milky Way, combining such observations with Galactic rotation curve information helps to distinguish multiple clouds along the LOS based on their high-resolution velocity information. In external galaxies, however, the problem appears more complex. Clouds superposition occurs at various scales (e.g. cores, fibers) and resolutions, and along the galactic heights \citep{Ballesteros_19}. Even though high-critical density line observations in external galaxies cannot by themselves resolve the issue of cloud superposition, they can mitigate the problem by ensuring that the emission originates from high-density regions rather than the accumulation of low densities along the LOS. Such observations have been recently conducted by \cite{MJ_23} who presented the first systematic extragalactic observations of N$_2$H$^{+}$(1--0) and HCN(1--0) in a wide range of dynamical conditions and star formation properties sampled across an entire galaxy disk.

It is beyond the scope of this paper to quantify biases in extragalactic studies, as it demands a more comprehensive analysis of involved factors. However, it is evident that properties derived solely from observed column densities carry potential biases, especially with recent advancements in extragalactic observations that capture individual cloud sizes \citep[e.g.][]{Schinnerer13,Faesi_18,Leroy21,Sun20}. 
\subsection{Caveats}
\label{sec:caveats}
Despite its great potential and strength, it is crucial to acknowledge the limitations in our study that require consideration. These aspects offer insights into the boundaries and potential constraints of our findings.

As mentioned earlier, one of the main limiting factors in our work lies within the input data. Given the observed patterns of APOGEE, the space is not uniformly observed. There are various gaps in the data both in the local neighbourhood and further outside. As a result, our map underestimates the dust distribution in some regions, the total number of clouds, and, therefore, the total dust mass of the Galaxy. This particularly affects the fourth quadrant, and near the Galactic centre at b=0, both of which are observed very sparsely.

Another limitation imposed on our results due to the incomplete input data is the final resolution. As explained in section \ref{sec:method}, the typical separation between the input stars sets the cell size for our model and affects the final resolution of the map. Therefore, given the sparsity in the observed APOGEE data, our final resolution is $\sim 100$ pc. This affects the clouds reported in our catalogue: our map is unable to resolve multiple small clouds that appear in close vicinity of one another; thus, some of our large clouds could contain unresolved substructures at various distances and densities within the cloud range. An example of this is around the Vela molecular cloud at X = -8 kpc, Y = -2 kpc, which includes multiple clouds in the map of \cite{Vergely_22} with better resolution but limited distances.

Another possible constraint of our work is in the clouds selected from our map for further analyses (Fig. \ref{fig:3D_all}, left). To avoid biases, we applied a strict cut to select clouds from our map (see section \ref{sec:features}). This results in missing potential low- to mid-density clouds that have values below or around three sigma above the mean of the Gaussian process. Therefore, the selected clouds in our map are incomplete for low to mid densities.

Apart from the catalogue of large molecular clouds in the Milky Way, which is carefully selected, we have also included our 3D density predictions with this publication. The users are however advised to use the map carefully and only with its predicted uncertainty to avoid biases and noisy outputs.
\section{Concluding remarks}
\label{sec:conclusion}

We have presented the most extended 3D dust map of the Milky Way to date and provided a catalogue of large molecular clouds in the Milky Way. The cloud properties in the catalogue are derived from the 3D map and avoid biases involved in plane-of-the-sky works. The catalogue delivers (non-kinematic) accurate distance estimates to high-density regains and contains their volume densities. Our map illustrates large cavities in the Galactic Plane, posing as potential targets for further studies and analysis.

Our 3D map sheds light on segments of the spiral arms; however, we do not observe clear arm patterns in our results. Using the volume densities derived from our map, we also studied the distribution of mass for different Galactocentric radii. We observe an overall decreasing trend as we approach the outer Galaxy, followed by multiple local peaks linked to known regions, such as the Molecular Ring, and segments of the spiral arms.

Additionally, our results provide insights into the extragalactic studies which focus on deriving properties of the clouds in star-forming regions. We demonstrate how the inferred properties of star-forming regions could be biased by using column density measurements, and suggest observations of molecular line emissions in critical high-density environments to minimise such biases. Further studies are required to quantify these biases.

The future of the 3D structure of the Milky Way relies on a uniform, near-infrared survey that covers the entire Galactic plane. This will soon be achieved by the upcoming SDSS-V data \citep{Kollmeier17}, which follows the legacy of APOGEE and is poised to achieve this ambitious goal.


\begin{acknowledgements}
\\The authors acknowledge Interstellar Institute's program "II6" and the Paris-Saclay University's Institut Pascal for hosting discussions that nourished the development of the ideas behind this work.\\
SRKh's goodbye letter to astronomy:\\
\\
\textit{Dear Astronomy*, \\
\\
From the day I started noticing the night sky, when grandma pointed to that bright sunset "star", when I was an astronaut in my dreams flying into the dark infinite beauty of space, I knew one thing was real: I fell in love with you. A love so deep, I gave you my all: from my mind and body, to leaving my family behind... I knew I had one goal: to get to discover you, to let you take me with you. \\
I did everything for you because that's what you do to someone who makes you feel as alive as you've made me feel, and I will always love you for it. But I can't love you excessively for much longer; this paper is all I have left to give. My curiosity still has a lot unanswered, my body can still handle the grind of sitting all day, but my mental health knows it's time to say goodbye. My soul covered with so many wounds doesn't leave me a choice. Wounds so deep I'll carry them for the rest of my life.\\
I know I will miss you every time I look at the beautiful forgotten night sky, but this is for the best; academia is a dark cloudy sky, and I don't want its darkness to put a shadow on our love. From the good and the bad, we have given each other all that we had, and we both know no matter what I do next, I will always be that kid whose eyes widen whenever she looks at the sunset "star".\\
\\
Love you always,\\
Sara Rezaei Kh.\\
\\
*inspired by and partially adopted from 'Dear Basketball', Kobe Bryant, 1978-2020\\
\url{https://en.m.wikipedia.org/wiki/Dear_Basketball}
}
\end{acknowledgements}
\bibliographystyle{aa}
\bibliography{3D_GP}

\begin{thebibliography}{86}
\expandafter\ifx\csname natexlab\endcsname\relax\def\natexlab#1{#1}\fi

\bibitem[{{Abolfathi} {et~al.}(2018){Abolfathi}, {Aguado},
  {et~al.}}]{Abolfathi17}
{Abolfathi}, B., {Aguado}, D.~S., {et~al.} 2018, \apjs, 235, 42

\bibitem[{{Ballesteros-Paredes} {et~al.}(2019){Ballesteros-Paredes},
  {Rom{\'a}n-Z{\'u}{\~n}iga}, {Salom{\'e}}, {Zamora-Avil{\'e}s}, \&
  {Jim{\'e}nez-Donaire}}]{Ballesteros_19}
{Ballesteros-Paredes}, J., {Rom{\'a}n-Z{\'u}{\~n}iga}, C., {Salom{\'e}}, Q.,
  {Zamora-Avil{\'e}s}, M., \& {Jim{\'e}nez-Donaire}, M.~J. 2019, \mnras, 490,
  2648

\bibitem[{{Bally} {et~al.}(1987){Bally}, {Langer}, {Stark}, \&
  {Wilson}}]{Bally_87}
{Bally}, J., {Langer}, W.~D., {Stark}, A.~A., \& {Wilson}, R.~W. 1987, \apjl,
  312, L45

\bibitem[{{Be{\v{s}}li{\'c}} {et~al.}(2021){Be{\v{s}}li{\'c}}, {Barnes},
  {Bigiel}, {Puschnig}, {Pety}, {Herrera Contreras}, {Leroy}, {Usero},
  {Schinnerer}, {Meidt}, {Emsellem}, {Hughes}, {Faesi}, {Kreckel}, {Belfiore},
  {Chevance}, {den Brok}, {Eibensteiner}, {Glover}, {Grasha},
  {Jimenez-Donaire}, {Klessen}, {Kruijssen}, {Liu}, {Pessa}, {Querejeta},
  {Rosolowsky}, {Saito}, {Santoro}, {Schruba}, {Sormani}, \&
  {Williams}}]{Beslic_21}
{Be{\v{s}}li{\'c}}, I., {Barnes}, A.~T., {Bigiel}, F., {et~al.} 2021, \mnras,
  506, 963

\bibitem[{{Binder} \& {Povich}(2018)}]{Binder_18}
{Binder}, B.~A. \& {Povich}, M.~S. 2018, \apj, 864, 136

\bibitem[{{Blanton} {et~al.}(2017){Blanton}, {Bershady}, {et~al.}}]{Blanton17}
{Blanton}, M.~R., {Bershady}, M.~A., {et~al.} 2017, \aj, 154, 28

\bibitem[{{Brand} \& {Blitz}(1993)}]{Brand93}
{Brand}, J. \& {Blitz}, L. 1993, \aap, 275, 67

\bibitem[{{Brown}(2021)}]{Brown21}
{Brown}, A. G.~A. 2021, \araa, 59, 59

\bibitem[{{Cahlon} {et~al.}(2024){Cahlon}, {Zucker}, {Goodman}, {Lada}, \&
  {Alves}}]{Cahlon_24}
{Cahlon}, S., {Zucker}, C., {Goodman}, A., {Lada}, C., \& {Alves}, J. 2024,
  \apj, 961, 153

\bibitem[{{Cantat-Gaudin} {et~al.}(2018){Cantat-Gaudin}, {Jordi},
  {et~al.}}]{Cantat18}
{Cantat-Gaudin}, T., {Jordi}, C., {et~al.} 2018, \aap, 618, A93

\bibitem[{{Chiappini} {et~al.}(2001){Chiappini}, {Matteucci}, \&
  {Romano}}]{Chiappini_01}
{Chiappini}, C., {Matteucci}, F., \& {Romano}, D. 2001, \apj, 554, 1044

\bibitem[{{Colombo} {et~al.}(2014){Colombo}, {Hughes}, {Schinnerer}, {Meidt},
  {Leroy}, {Pety}, {Dobbs}, {Garc{\'\i}a-Burillo}, {Dumas}, {Thompson},
  {Schuster}, \& {Kramer}}]{Colombo_14}
{Colombo}, D., {Hughes}, A., {Schinnerer}, E., {et~al.} 2014, \apj, 784, 3

\bibitem[{{Dame} {et~al.}(1986){Dame}, {Elmegreen}, {Cohen}, \&
  {Thaddeus}}]{Dame_86}
{Dame}, T.~M., {Elmegreen}, B.~G., {Cohen}, R.~S., \& {Thaddeus}, P. 1986,
  \apj, 305, 892

\bibitem[{{Dame} {et~al.}(2001){Dame}, {Hartmann}, {et~al.}}]{Dame01}
{Dame}, T.~M., {Hartmann}, D., {et~al.} 2001, \apj, 547, 792

\bibitem[{{Dame} \& {Lada}(2023)}]{Dame_23}
{Dame}, T.~M. \& {Lada}, C.~J. 2023, \apj, 944, 197

\bibitem[{{Darling} {et~al.}(2023){Darling}, {Paine}, {Reid}, {Menten},
  {Sakai}, \& {Ghez}}]{Darling_23}
{Darling}, J., {Paine}, J., {Reid}, M.~J., {et~al.} 2023, \apj, 955, 117

\bibitem[{{Donovan Meyer} {et~al.}(2013){Donovan Meyer}, {Koda}, {Momose},
  {Mooney}, {Egusa}, {Carty}, {Kennicutt}, {Kuno}, {Rebolledo}, {Sawada},
  {Scoville}, \& {Wong}}]{Donovan_13}
{Donovan Meyer}, J., {Koda}, J., {Momose}, R., {et~al.} 2013, \apj, 772, 107

\bibitem[{{Draine}(2009)}]{Draine_09}
{Draine}, B.~T. 2009, in Astronomical Society of the Pacific Conference Series,
  Vol. 414, Cosmic Dust - Near and Far, ed. T.~{Henning}, E.~{Gr{\"u}n}, \&
  J.~{Steinacker}, 453

\bibitem[{{Drimmel} \& {Spergel}(2001)}]{Drimmel01}
{Drimmel}, R. \& {Spergel}, D.~N. 2001, \apj, 556, 181

\bibitem[{{Edenhofer} {et~al.}(2023){Edenhofer}, {Zucker}, {Frank}, {Saydjari},
  {Speagle}, {Finkbeiner}, \& {En{\ss}lin}}]{Eden_23}
{Edenhofer}, G., {Zucker}, C., {Frank}, P., {et~al.} 2023, arXiv e-prints,
  arXiv:2308.01295

\bibitem[{{Eisenstein} {et~al.}(2011){Eisenstein}, {Weinberg},
  {et~al.}}]{Eisenstein11}
{Eisenstein}, D.~J., {Weinberg}, D.~H., {et~al.} 2011, \aj, 142, 72

\bibitem[{{Elia} {et~al.}(2017){Elia}, {Molinari}, {Schisano}, {Pestalozzi},
  {Pezzuto}, {Merello}, {Noriega-Crespo}, {Moore}, {Russeil}, {Mottram},
  {Paladini}, {Strafella}, {Benedettini}, {Bernard}, {Di Giorgio}, {Eden},
  {Fukui}, {Plume}, {Bally}, {Martin}, {Ragan}, {Jaffa}, {Motte}, {Olmi},
  {Schneider}, {Testi}, {Wyrowski}, {Zavagno}, {Calzoletti}, {Faustini},
  {Natoli}, {Palmeirim}, {Piacentini}, {Piazzo}, {Pilbratt}, {Polychroni},
  {Baldeschi}, {Beltr{\'a}n}, {Billot}, {Cambr{\'e}sy}, {Cesaroni},
  {Garc{\'\i}a-Lario}, {Hoare}, {Huang}, {Joncas}, {Liu}, {Maiolo}, {Marsh},
  {Maruccia}, {M{\`e}ge}, {Peretto}, {Rygl}, {Schilke}, {Thompson},
  {Traficante}, {Umana}, {Veneziani}, {Ward-Thompson}, {Whitworth}, {Arab},
  {Bandieramonte}, {Becciani}, {Brescia}, {Buemi}, {Bufano}, {Butora},
  {Cavuoti}, {Costa}, {Fiorellino}, {Hajnal}, {Hayakawa}, {Kacsuk}, {Leto}, {Li
  Causi}, {Marchili}, {Martinavarro-Armengol}, {Mercurio}, {Molinaro},
  {Riccio}, {Sano}, {Sciacca}, {Tachihara}, {Torii}, {Trigilio}, {Vitello}, \&
  {Yamamoto}}]{Elia_17}
{Elia}, D., {Molinari}, S., {Schisano}, E., {et~al.} 2017, \mnras, 471, 100

\bibitem[{{Elia} {et~al.}(2022){Elia}, {Molinari}, {Schisano}, {Soler},
  {Merello}, {Russeil}, {Veneziani}, {Zavagno}, {Noriega-Crespo}, {Olmi},
  {Benedettini}, {Hennebelle}, {Klessen}, {Leurini}, {Paladini}, {Pezzuto},
  {Traficante}, {Eden}, {Martin}, {Sormani}, {Coletta}, {Colman}, {Plume},
  {Maruccia}, {Mininni}, \& {Liu}}]{Elia_22}
{Elia}, D., {Molinari}, S., {Schisano}, E., {et~al.} 2022, \apj, 941, 162

\bibitem[{{Evans} {et~al.}(2020){Evans}, {Kim}, {Wu}, {Chao}, {Heyer}, {Liu},
  {Nguyen-Lu'o'ng}, \& {Kauffmann}}]{Evans_20}
{Evans}, Neal~J., I., {Kim}, K.-T., {Wu}, J., {et~al.} 2020, \apj, 894, 103

\bibitem[{{Faesi} {et~al.}(2018){Faesi}, {Lada}, \& {Forbrich}}]{Faesi_18}
{Faesi}, C.~M., {Lada}, C.~J., \& {Forbrich}, J. 2018, \apj, 857, 19

\bibitem[{{Freeman} {et~al.}(2017){Freeman}, {Rosolowsky}, {Kruijssen},
  {Bastian}, \& {Adamo}}]{Freeman_17}
{Freeman}, P., {Rosolowsky}, E., {Kruijssen}, J.~M.~D., {Bastian}, N., \&
  {Adamo}, A. 2017, \mnras, 468, 1769

\bibitem[{{Gaia Collaboration} {et~al.}(2016){Gaia Collaboration}, {Prusti},
  {et~al.}}]{Gaia_collaboration16}
{Gaia Collaboration}, {Prusti}, T., {et~al.} 2016, \aap, 595, A1

\bibitem[{{Gallagher} {et~al.}(2018){Gallagher}, {Leroy}, {Bigiel}, {Cormier},
  {Jim{\'e}nez-Donaire}, {Ostriker}, {Usero}, {Bolatto}, {Garc{\'\i}a-Burillo},
  {Hughes}, {Kepley}, {Krumholz}, {Meidt}, {Meier}, {Murphy}, {Pety},
  {Rosolowsky}, {Schinnerer}, {Schruba}, \& {Walter}}]{Gallagher_18}
{Gallagher}, M.~J., {Leroy}, A.~K., {Bigiel}, F., {et~al.} 2018, \apj, 858, 90

\bibitem[{{GRAVITY Collaboration} {et~al.}(2021){GRAVITY Collaboration},
  {Abuter}, {Amorim}, {Baub{\"o}ck}, {Berger}, {Bonnet}, {Brandner},
  {Cl{\'e}net}, {Davies}, {de Zeeuw}, {Dexter}, {Dallilar}, {Drescher},
  {Eckart}, {Eisenhauer}, {F{\"o}rster Schreiber}, {Garcia}, {Gao}, {Gendron},
  {Genzel}, {Gillessen}, {Habibi}, {Haubois}, {Hei{\ss}el}, {Henning},
  {Hippler}, {Horrobin}, {Jim{\'e}nez-Rosales}, {Jochum}, {Jocou}, {Kaufer},
  {Kervella}, {Lacour}, {Lapeyr{\`e}re}, {Le Bouquin}, {L{\'e}na}, {Lutz},
  {Nowak}, {Ott}, {Paumard}, {Perraut}, {Perrin}, {Pfuhl}, {Rabien},
  {Rodr{\'\i}guez-Coira}, {Shangguan}, {Shimizu}, {Scheithauer}, {Stadler},
  {Straub}, {Straubmeier}, {Sturm}, {Tacconi}, {Vincent}, {von Fellenberg},
  {Waisberg}, {Widmann}, {Wieprecht}, {Wiezorrek}, {Woillez}, {Yazici},
  {Young}, \& {Zins}}]{Gravity_21}
{GRAVITY Collaboration}, {Abuter}, R., {Amorim}, A., {et~al.} 2021, \aap, 647,
  A59

\bibitem[{{Green} {et~al.}(2019{\natexlab{a}}){Green}, {Schlafly}, {Zucker},
  {Speagle}, \& {Finkbeiner}}]{Green_19}
{Green}, G.~M., {Schlafly}, E., {Zucker}, C., {Speagle}, J.~S., \&
  {Finkbeiner}, D. 2019{\natexlab{a}}, \apj, 887, 93

\bibitem[{{Green} {et~al.}(2019{\natexlab{b}}){Green}, {Schlafly},
  {et~al.}}]{Green19}
{Green}, G.~M., {Schlafly}, E., {et~al.} 2019{\natexlab{b}}, \apj, 887, 93

\bibitem[{{Gro\ss{}schedl} {et~al.}(2018){Gro\ss{}schedl}, {Alves, Jo\~ao},
  {Meingast, Stefan}, {Ackerl, Christine}, {Ascenso, Joana}, {Bouy, Herv\'e},
  {Burkert, Andreas}, {Forbrich, Jan}, {F\"urnkranz, Verena}, {Goodman,
  Alyssa}, {Hacar, \'Alvaro}, {Herbst-Kiss, Gabor}, {Lada, Charles J.},
  {Larreina, Irati}, {Leschinski, Kieran}, {Lombardi, Marco}, {Moitinho,
  Andr\'e}, {Mortimer, Daniel}, \& {Zari, Eleonora}}]{Grossscheld18}
{Gro\ss{}schedl}, J.~E., {Alves, Jo\~ao}, {Meingast, Stefan}, {et~al.} 2018,
  A\&A, 619, A106

\bibitem[{{Henshaw} {et~al.}(2023){Henshaw}, {Barnes}, {Battersby}, {Ginsburg},
  {Sormani}, \& {Walker}}]{Henshaw_23}
{Henshaw}, J.~D., {Barnes}, A.~T., {Battersby}, C., {et~al.} 2023, in
  Astronomical Society of the Pacific Conference Series, Vol. 534, Protostars
  and Planets VII, ed. S.~{Inutsuka}, Y.~{Aikawa}, T.~{Muto}, K.~{Tomida}, \&
  M.~{Tamura}, 83

\bibitem[{{Hogg} {et~al.}(2019){Hogg}, {Eilers}, {et~al.}}]{Hogg19}
{Hogg}, D.~W., {Eilers}, A.-C., {et~al.} 2019, \aj, 158, 147

\bibitem[{{Jim{\'e}nez-Donaire} {et~al.}(2019){Jim{\'e}nez-Donaire}, {Bigiel},
  {Leroy}, {Usero}, {Cormier}, {Puschnig}, {Gallagher}, {Kepley}, {Bolatto},
  {Garc{\'\i}a-Burillo}, {Hughes}, {Kramer}, {Pety}, {Schinnerer}, {Schruba},
  {Schuster}, \& {Walter}}]{MJ_19}
{Jim{\'e}nez-Donaire}, M.~J., {Bigiel}, F., {Leroy}, A.~K., {et~al.} 2019,
  \apj, 880, 127

\bibitem[{{Jim{\'e}nez-Donaire} {et~al.}(2023){Jim{\'e}nez-Donaire}, {Usero},
  {Be{\v{s}}li{\'c}}, {Tafalla}, {Chac{\'o}n-Tanarro}, {Salom{\'e}},
  {Eibensteiner}, {Garc{\'\i}a-Rodr{\'\i}guez}, {Hacar}, {Barnes}, {Bigiel},
  {Chevance}, {Colombo}, {Dale}, {Davis}, {Glover}, {Kauffmann}, {Klessen},
  {Leroy}, {Neumann}, {Pan}, {Pety}, {Querejeta}, {Saito}, {Schinnerer},
  {Stuber}, \& {Williams}}]{MJ_23}
{Jim{\'e}nez-Donaire}, M.~J., {Usero}, A., {Be{\v{s}}li{\'c}}, I., {et~al.}
  2023, \aap, 676, L11

\bibitem[{{J{\"o}nsson} {et~al.}(2020){J{\"o}nsson}, {Holtzman}, {Allende
  Prieto}, {Cunha}, {Garc{\'\i}a-Hern{\'a}ndez}, {Hasselquist}, {Masseron},
  {Osorio}, {Shetrone}, {Smith}, {Stringfellow}, {Bizyaev}, {Edvardsson},
  {Majewski}, {M{\'e}sz{\'a}ros}, {Souto}, {Zamora}, {Beaton}, {Bovy}, {Donor},
  {Pinsonneault}, {Poovelil}, \& {Sobeck}}]{APOGEE16}
{J{\"o}nsson}, H., {Holtzman}, J.~A., {Allende Prieto}, C., {et~al.} 2020, \aj,
  160, 120

\bibitem[{{Kainulainen} {et~al.}(2022){Kainulainen}, {Rezaei}, {Spilker}, \&
  {Orkisz}}]{Kainulainen22}
{Kainulainen}, J., {Rezaei}, K.~S., {Spilker}, A., \& {Orkisz}, J. 2022, \aap,
  659, L6

\bibitem[{{Kalberla} \& {Kerp}(2009)}]{Kalberla09}
{Kalberla}, P. M.~W. \& {Kerp}, J. 2009, \araa, 47, 27

\bibitem[{{Kauffmann} {et~al.}(2017){Kauffmann}, {Goldsmith}, {Melnick},
  {Tolls}, {Guzman}, \& {Menten}}]{Kauffmann_17}
{Kauffmann}, J., {Goldsmith}, P.~F., {Melnick}, G., {et~al.} 2017, \aap, 605,
  L5

\bibitem[{{Kawamura} {et~al.}(2009){Kawamura}, {Mizuno}, {Minamidani},
  {Filipovi{\'c}}, {Staveley-Smith}, {Kim}, {Mizuno}, {Onishi}, {Mizuno}, \&
  {Fukui}}]{Kawamura_09}
{Kawamura}, A., {Mizuno}, Y., {Minamidani}, T., {et~al.} 2009, \apjs, 184, 1

\bibitem[{{Kennicutt} \& {Evans}(2012)}]{Kennicutt12}
{Kennicutt}, R.~C. \& {Evans}, N.~J. 2012, \araa, 50, 531

\bibitem[{{Kollmeier} {et~al.}(2017)}]{Kollmeier17}
{Kollmeier}, J.~A. {et~al.} 2017, arXiv e-prints, 1711.03234

\bibitem[{{Krumholz} \& {McKee}(2005)}]{Krumholz_05}
{Krumholz}, M.~R. \& {McKee}, C.~F. 2005, \apj, 630, 250

\bibitem[{{Krumholz} {et~al.}(2018)}]{Krumholz18}
{Krumholz}, M.~R. {et~al.} 2018, \mnras, 477, 2716

\bibitem[{{Kuno} {et~al.}(2007){Kuno}, {Sato}, {Nakanishi}, {Hirota}, {Tosaki},
  {Shioya}, {Sorai}, {Nakai}, {Nishiyama}, \& {Vila-Vilar{\'o}}}]{Kuno07}
{Kuno}, N., {Sato}, N., {Nakanishi}, H., {et~al.} 2007, \pasj, 59, 117

\bibitem[{{Lee} {et~al.}(2016){Lee}, {Miville-Desch{\^e}nes}, \&
  {Murray}}]{Lee_16}
{Lee}, E.~J., {Miville-Desch{\^e}nes}, M.-A., \& {Murray}, N.~W. 2016, \apj,
  833, 229

\bibitem[{{Lee} {et~al.}(2023){Lee}, {Sandstrom}, {Leroy}, {Thilker},
  {Schinnerer}, {Rosolowsky}, {Larson}, {Egorov}, {Williams}, {Schmidt},
  {Emsellem}, {Anand}, {Barnes}, {Belfiore}, {Be{\v{s}}li{\'c}}, {Bigiel},
  {Blanc}, {Bolatto}, {Boquien}, {den Brok}, {Cao}, {Chandar}, {Chastenet},
  {Chevance}, {Chiang}, {Congiu}, {Dale}, {Deger}, {Eibensteiner}, {Faesi},
  {Glover}, {Grasha}, {Groves}, {Hassani}, {Henny}, {Henshaw}, {Hoyer},
  {Hughes}, {Jeffreson}, {Jim{\'e}nez-Donaire}, {Kim}, {Kim}, {Klessen},
  {Koch}, {Kreckel}, {Kruijssen}, {Li}, {Liu}, {Lopez}, {Maschmann}, {Chen},
  {Meidt}, {Murphy}, {Neumann}, {Neumayer}, {Pan}, {Pessa}, {Pety},
  {Querejeta}, {Pinna}, {Rodr{\'\i}guez}, {Saito}, {S{\'a}nchez-Bl{\'a}zquez},
  {Santoro}, {Sardone}, {Smith}, {Sormani}, {Scheuermann}, {Stuber}, {Sutter},
  {Sun}, {Teng}, {Tre{\ss}}, {Usero}, {Watkins}, {Whitmore}, \&
  {Razza}}]{Lee_23}
{Lee}, J.~C., {Sandstrom}, K.~M., {Leroy}, A.~K., {et~al.} 2023, \apjl, 944,
  L17

\bibitem[{{Leike} {et~al.}(2020){Leike}, {Glatzle}, \& {En{\ss}lin}}]{Leike_20}
{Leike}, R.~H., {Glatzle}, M., \& {En{\ss}lin}, T.~A. 2020, \aap, 639, A138

\bibitem[{{Leroy} {et~al.}(2021){Leroy}, {Schinnerer}, {et~al.}}]{Leroy21}
{Leroy}, A.~K., {Schinnerer}, E., {et~al.} 2021, \apjs, 257, 43

\bibitem[{{Leung} {et~al.}(2023){Leung}, {Bovy}, {Mackereth}, {Hunt}, {Lane},
  \& {Wilson}}]{Leung_23}
{Leung}, H.~W., {Bovy}, J., {Mackereth}, J.~T., {et~al.} 2023, \mnras, 519, 948

\bibitem[{{Majewski} {et~al.}(2017){Majewski}, {Schiavon},
  {et~al.}}]{Majewski17}
{Majewski}, S.~R., {Schiavon}, R.~P., {et~al.} 2017, \aj, 154, 94

\bibitem[{{Majewski} {et~al.}(2011){Majewski}, {Zasowski},
  {et~al.}}]{Majewski11}
{Majewski}, S.~R., {Zasowski}, G., {et~al.} 2011, \apj, 739, 25

\bibitem[{{Marshall} {et~al.}(2006){Marshall}, {Robin}, {et~al.}}]{Marshall06}
{Marshall}, D.~J., {Robin}, A.~C., {et~al.} 2006, \aap, 453, 635

\bibitem[{{Miville-Desch{\^e}nes} {et~al.}(2017)}]{Miville17}
{Miville-Desch{\^e}nes}, M.-A. {et~al.} 2017, \apj, 834, 57

\bibitem[{{Morgan}(1955)}]{Morgan55}
{Morgan}, W.~W. 1955, Scientific American, 192, 42

\bibitem[{{Oort} \& {Muller}(1952)}]{Oort52}
{Oort}, J.~H. \& {Muller}, C.~A. 1952, Monthly Notes of the Astronomical
  Society of South Africa, 11, 65

\bibitem[{{Ou} {et~al.}(2023){Ou}, {Eilers}, {Necib}, \& {Frebel}}]{Ou_23}
{Ou}, X., {Eilers}, A.-C., {Necib}, L., \& {Frebel}, A. 2023, arXiv e-prints,
  arXiv:2303.12838

\bibitem[{{Padoan} {et~al.}(2014){Padoan}, {Federrath}, {et~al.}}]{Padoan14}
{Padoan}, P., {Federrath}, C., {et~al.} 2014, in Protostars and Planets VI,
  77--100

\bibitem[{{Pety} {et~al.}(2017){Pety}, {Guzm{\'a}n}, {Orkisz}, {Liszt},
  {Gerin}, {Bron}, {Bardeau}, {Goicoechea}, {Gratier}, {Le Petit}, {Levrier},
  {{\"O}berg}, {Roueff}, \& {Sievers}}]{Pety_17}
{Pety}, J., {Guzm{\'a}n}, V.~V., {Orkisz}, J.~H., {et~al.} 2017, \aap, 599, A98

\bibitem[{{Poggio} {et~al.}(2020){Poggio}, {Drimmel}, {Andrae}, {Bailer-Jones},
  {Fouesneau}, {Lattanzi}, {Smart}, \& {Spagna}}]{Poggio_20}
{Poggio}, E., {Drimmel}, R., {Andrae}, R., {et~al.} 2020, Nature Astronomy, 4,
  590

\bibitem[{{Querejeta} {et~al.}(2019){Querejeta}, {Schinnerer}, {Schruba},
  {Murphy}, {Meidt}, {Usero}, {Leroy}, {Pety}, {Bigiel}, {Chevance}, {Faesi},
  {Gallagher}, {Garc{\'\i}a-Burillo}, {Glover}, {Hygate},
  {Jim{\'e}nez-Donaire}, {Kruijssen}, {Momjian}, {Rosolowsky}, \&
  {Utomo}}]{Querejeta_19}
{Querejeta}, M., {Schinnerer}, E., {Schruba}, A., {et~al.} 2019, \aap, 625, A19

\bibitem[{{Reid} {et~al.}(2016){Reid}, {Dame}, {Menten}, \&
  {Brunthaler}}]{BESSEL_16}
{Reid}, M.~J., {Dame}, T.~M., {Menten}, K.~M., \& {Brunthaler}, A. 2016, \apj,
  823, 77

\bibitem[{{Reid} {et~al.}(2019){Reid}, {Menten}, {et~al.}}]{Reid19}
{Reid}, M.~J., {Menten}, K.~M., {et~al.} 2019, \apj, 885, 131

\bibitem[{{Rezaei Kh.} {et~al.}(2017){Rezaei Kh.}, {Bailer-Jones}, {Hanson}, \&
  {Fouesneau}}]{Rezaei_Kh_17}
{Rezaei Kh.}, S., {Bailer-Jones}, C.~A.~L., {Hanson}, R.~J., \& {Fouesneau}, M.
  2017, \aap, 598, A125

\bibitem[{{Rezaei Kh.} {et~al.}(2018{\natexlab{a}}){Rezaei Kh.},
  {Bailer-Jones}, {Schlafly}, \& {Fouesneau}}]{Rezaei_Kh_18a}
{Rezaei Kh.}, S., {Bailer-Jones}, C.~A.~L., {Schlafly}, E.~F., \& {Fouesneau},
  M. 2018{\natexlab{a}}, \aap, 616, A44

\bibitem[{{Rezaei Kh.} {et~al.}(2020){Rezaei Kh.}, {Bailer-Jones}, {Soler}, \&
  {Zari}}]{Rezaei_Kh_20}
{Rezaei Kh.}, S., {Bailer-Jones}, C. A.~L., {Soler}, J.~D., \& {Zari}, E. 2020,
  \aap, 643, A151

\bibitem[{{Rezaei Kh.} {et~al.}(2018{\natexlab{b}}){Rezaei Kh.}, {Bailer-Jones,
  Coryn A. L.}, {Hogg, David W.}, \& {Schultheis, Mathias}}]{Rezaei_Kh_18b}
{Rezaei Kh.}, S., {Bailer-Jones, Coryn A. L.}, {Hogg, David W.}, \&
  {Schultheis, Mathias}. 2018{\natexlab{b}}, A\&A, 618, A168

\bibitem[{{Rezaei Kh.} \& {Kainulainen}(2022)}]{Rezaei_Kh22}
{Rezaei Kh.}, S. \& {Kainulainen}, J. 2022, \apjl, 930, L22

\bibitem[{{Rodgers} {et~al.}(1960){Rodgers}, {Campbell}, \&
  {Whiteoak}}]{Rodgers_60}
{Rodgers}, A.~W., {Campbell}, C.~T., \& {Whiteoak}, J.~B. 1960, \mnras, 121,
  103

\bibitem[{{Roman-Duval} {et~al.}(2009){Roman-Duval}, {Jackson}, {Heyer},
  {Johnson}, {Rathborne}, {Shah}, \& {Simon}}]{Roman-Duval_09}
{Roman-Duval}, J., {Jackson}, J.~M., {Heyer}, M., {et~al.} 2009, \apj, 699,
  1153

\bibitem[{{Roman-Duval} {et~al.}(2010){Roman-Duval}, {Jackson},
  {et~al.}}]{Roman10}
{Roman-Duval}, J., {Jackson}, J.~M., {et~al.} 2010, \apj, 723, 492

\bibitem[{{Romero-G{\'o}mez} {et~al.}(2019){Romero-G{\'o}mez}, {Antoja},
  {Figueras}, {Mateu}, {Aguilar}, {Castro-Ginard}, {Carrasco}, \&
  {Masana}}]{Romero19}
{Romero-G{\'o}mez}, M., {Antoja}, T., {Figueras}, F., {et~al.} 2019, in
  Highlights on Spanish Astrophysics X, ed. B.~{Montesinos}, A.~{Asensio
  Ramos}, F.~{Buitrago}, R.~{Sch{\"o}del}, E.~{Villaver}, S.~{P{\'e}rez-Hoyos},
  \& I.~{Ord{\'o}{\~n}ez-Etxeberria}, 386--391

\bibitem[{{Russeil}(2003)}]{Russeil03}
{Russeil}, D. 2003, \aap, 397, 133

\bibitem[{{Schinnerer} {et~al.}(2013){Schinnerer}, {Meidt},
  {et~al.}}]{Schinnerer13}
{Schinnerer}, E., {Meidt}, S.~E., {et~al.} 2013, \apj, 779, 42

\bibitem[{{Skrutskie} {et~al.}(2006){Skrutskie}, {Cutri},
  {et~al.}}]{Skrutskie06}
{Skrutskie}, M.~F., {Cutri}, R.~M., {et~al.} 2006, \aj, 131, 1163

\bibitem[{{Sun} {et~al.}(2022){Sun}, {Leroy}, {Rosolowsky}, {Hughes},
  {Schinnerer}, {Schruba}, {Koch}, {Blanc}, {Chiang}, {Groves}, {Liu}, {Meidt},
  {Pan}, {Pety}, {Querejeta}, {Saito}, {Sandstrom}, {Sardone}, {Usero},
  {Utomo}, {Williams}, {Barnes}, {Benincasa}, {Bigiel}, {Bolatto}, {Boquien},
  {Chevance}, {Dale}, {Deger}, {Emsellem}, {Glover}, {Grasha}, {Henshaw},
  {Klessen}, {Kreckel}, {Kruijssen}, {Ostriker}, \& {Thilker}}]{Sun_22}
{Sun}, J., {Leroy}, A.~K., {Rosolowsky}, E., {et~al.} 2022, \aj, 164, 43

\bibitem[{{Sun} {et~al.}(2020){Sun}, {Leroy}, {et~al.}}]{Sun20}
{Sun}, J., {Leroy}, A.~K., {et~al.} 2020, \apjl, 901, L8

\bibitem[{{Tafalla} {et~al.}(2021){Tafalla}, {Usero}, \& {Hacar}}]{Tafalla_21}
{Tafalla}, M., {Usero}, A., \& {Hacar}, A. 2021, \aap, 646, A97

\bibitem[{{Tafalla} {et~al.}(2023){Tafalla}, {Usero}, \& {Hacar}}]{Tafalla_23}
{Tafalla}, M., {Usero}, A., \& {Hacar}, A. 2023, \aap, 679, A112

\bibitem[{{van de Hulst} {et~al.}(1954){van de Hulst}, {Muller},
  {et~al.}}]{Hulst54}
{van de Hulst}, H.~C., {Muller}, C.~A., {et~al.} 1954, BAN, 12, 117

\bibitem[{{Vergely} {et~al.}(2022){Vergely}, {Lallement}, \&
  {Cox}}]{Vergely_22}
{Vergely}, J.~L., {Lallement}, R., \& {Cox}, N.~L.~J. 2022, \aap, 664, A174

\bibitem[{{Westerhout}(1958)}]{Westerhout_58}
{Westerhout}, G. 1958, \bain, 14, 215

\bibitem[{{Wright} {et~al.}(2010){Wright}, {Eisenhardt}, {et~al.}}]{Wright10}
{Wright}, E.~L., {Eisenhardt}, P.~R.~M., {et~al.} 2010, \aj, 140, 1868

\bibitem[{{Zasowski} {et~al.}(2013){Zasowski}, {Johnson},
  {et~al.}}]{Zasowski13}
{Zasowski}, G., {Johnson}, J.~A., {et~al.} 2013, \aj, 146, 81

\bibitem[{{Zucker} {et~al.}(2021){Zucker}, {Goodman}, {et~al.}}]{Zucker21}
{Zucker}, C., {Goodman}, A., {et~al.} 2021, \apj, 919, 35

\end{thebibliography}


\clearpage
\onecolumn
\begin{center}
\begin{longtable}{|l|l|l|l|l|l|l|l|l|}
\caption{Catalogue of large molecular clouds in the Galactic plane} \label{tab:clouds} \\

\hline \multicolumn{1}{|c|}{\textbf{Cloud}} & \multicolumn{1}{c|}{\textbf{l\_centre}} & \multicolumn{1}{c|}{\textbf{b\_centre}} & \multicolumn{1}{c|}{\textbf{d\_centre}} & \multicolumn{1}{c|}{\textbf{$\sigma_{d}$}} & \multicolumn{1}{c|}{\textbf{mean\_radius}} & \multicolumn{1}{c|}{\textbf{mean\_density}} & \multicolumn{1}{c|}{\textbf{density\_sd}} & {\textbf{Arm / }}\\ 
\multicolumn{1}{|c|}{\textbf{ID}} & \multicolumn{1}{c|}{[deg]} & \multicolumn{1}{c|}{[deg]} & \multicolumn{1}{c|}{[kpc]} & \multicolumn{1}{c|}{[kpc]} & \multicolumn{1}{c|}{[pc]} & \multicolumn{1}{c|}{[cm$^{-3}$]} & \multicolumn{1}{c|}{[cm$^{-3}$]} & {Association}\\ \hline 
\endfirsthead

\multicolumn{8}{c}%
{{\bfseries \tablename\ \thetable{} -- continued from previous page}} \\
\hline \multicolumn{1}{|c|}{\textbf{Cloud}} & \multicolumn{1}{c|}{\textbf{l\_centre}} & \multicolumn{1}{c|}{\textbf{b\_centre}} & \multicolumn{1}{c|}{\textbf{d\_centre}} & \multicolumn{1}{c|}{\textbf{d\_sd}} & \multicolumn{1}{c|}{\textbf{mean\_radius}} & \multicolumn{1}{c|}{\textbf{mean\_density}} & \multicolumn{1}{c|}{\textbf{density\_sd}} & {\textbf{Arm / }}\\ 
\multicolumn{1}{|c|}{\textbf{ID}} & \multicolumn{1}{c|}{[deg]} & \multicolumn{1}{c|}{[deg]} & \multicolumn{1}{c|}{[kpc]} & \multicolumn{1}{c|}{[kpc]} & \multicolumn{1}{c|}{[pc]} & \multicolumn{1}{c|}{[cm$^{-3}$]} & \multicolumn{1}{c|}{[cm$^{-3}$]} & {Association}\\ \hline 
\endhead

\hline \multicolumn{9}{|r|}{{Continued on next page}} \\ \hline
\endfoot

\midrule
\multicolumn{9}{l}{\parbox{\dimexpr\textwidth-2\tabcolsep}{%
\hspace{0.2cm} The columns are: cloud ID, Galactic coordinates of the centre of the cloud, distance to the centre of the cloud from the Sun, one sigma distance uncertainty of the centre of the cloud, mean radius and mean density of the cloud, standard deviation from the mean density, and corresponding known associations for the clouds. A cloud with a mean radius/density of NA indicates that only one pixel of the cloud has densities above the molecular threshold of 100 cm$^{-3}$; therefore, it is too small for our map to resolve its size.\\ 
The associations with spiral arms and known clouds are determined in multiple ways as follows:\\
1. Associated with spiral arms using masers in \cite{Reid19}\\
2. Having a corresponding cloud in \cite{Dame_86}, as explained in section \ref{sec:features} and Fig. \ref{fig:los}\\
3. Having a corresponding cloud in \cite{Westerhout_58}, \cite{Rodgers_60}, or \cite{Binder_18}
}}
\endlastfoot

    \hline
        \textbf{1} & 0.03 & 4.49 & 9.59 & 0.07 & 131 & 150.21 & 32.75 & \\ \hline
        \textbf{2} & 0.49 & -2.99 & 9.97 & 0.17 & 135 & 119.93 & 17.44 & \\ \hline
        \textbf{3} & 0.89 & -0.99 & 5.60 & 0.02 & 225 & 147.87 & 34.73 & \\ \hline
        \textbf{4} & 1.25 & -6.75 & 6.38 & 0.04 & 104 & 148.45 & 24.24 & \\ \hline
        \textbf{5} & 1.99 & 1.09 & 4.98 & 0.01 & 197 & 191.39 & 63.68 & Nor/Out$^{1}$ \\ \hline
        \textbf{6} & 2.39 & 5.10 & 8.44 & 0.09 & 85 & 134.36 & 17.83 & \\ \hline
        \textbf{7} & 3.95 & -2.46 & 8.73 & 0.94 & NA & 115.71 & NA & \\ \hline
        \textbf{8} & 4.01 & 2.51 & 7.63 & 0.10 & 152 & 133.21 & 18.39 & \\ \hline
        \textbf{9} & 4.15 & 1.41 & 6.08 & 0.20 & 53 & 109.48 & 4.15 & \\ \hline
        \textbf{10} & 5.80 & -3.23 & 6.66 & 0.04 & 93 & 149.50 & 29.40 & \\ \hline
        \textbf{11} & 5.93 & -2.49 & 9.67 & 0.04 & 285 & 141.94 & 42.26 & \\ \hline
        \textbf{12} & 6.15 & 0.00 & 2.17 & 0.01 & 163 & 147.65 & 26.13 & Sct/Cen$^{1}$ \\ \hline
        \textbf{13} & 7.14 & 5.39 & 7.98 & 0.16 & 81 & 114.48 & 12.80 & \\ \hline
        \textbf{14} & 7.36 & 2.92 & 9.51 & 0.03 & 195 & 157.09 & 39.26 & \\ \hline
        \textbf{15} & 7.81 & -0.67 & 3.66 & 0.01 & 187 & 135.55 & 29.96 & Nor/Out$^{1}$ \\ \hline
        \textbf{16} & 7.90 & 3.10 & 6.93 & 0.07 & 105 & 129.11 & 21.57 & \\ \hline
        \textbf{17} & 9.67 & -2.50 & 4.68 & 0.03 & 129 & 118.36 & 14.57 & \\ \hline
        \textbf{18} & 9.70 & -2.68 & 8.02 & 0.96 & 12 & 102.22 & 0.08 & \\ \hline
        \textbf{19} & 14.38 & 2.22 & 9.69 & 0.06 & 185 & 147.11 & 34.77 & \\ \hline
        \textbf{20} & 20.22 & 0.00 & 3.04 & 0.09 & 58 & 111.94 & 4.68 & Sct$^{2}$ \\ \hline
        \textbf{21} & 20.32 & 0.00 & 7.00 & 0.03 & 170 & 136.75 & 24.97 & \\ \hline
        \textbf{22} & 23.11 & -4.25 & 5.06 & 0.11 & 50 & 113.01 & 9.20 & Nor/Out$^{1}$ \\ \hline
        \textbf{23} & 23.36 & -6.29 & 6.85 & 0.38 & 65 & 111.13 & 0.87 & \\ \hline
        \textbf{24} & 26.08 & -7.04 & 6.12 & 0.09 & 59 & 122.72 & 13.80 & \\ \hline
        \textbf{25} & 26.81 & 0.00 & 1.21 & 0.01 & 99 & 118.29 & 16.20 & Sgr/Car$^{1,2}$ \\ \hline
        \textbf{26} & 28.03 & 0.00 & 10.24 & 0.05 & 325 & 113.69 & 9.55 & Sct/Cen$^{1,2}$ \\ \hline
        \textbf{27} & 28.76 & -2.83 & 7.60 & 0.20 & 86 & 107.02 & 4.16 & \\ \hline
        \textbf{28} & 29.36 & 0.00 & 3.67 & 0.70 & NA & 102.72 & NA & Sct/Cen$^{1,2}$ \\ \hline
        \textbf{29} & 40.07 & 0.00 & 2.06 & 0.01 & 128 & 116.28 & 9.70 & Sgr/Car$^{1,2}$ \\ \hline
        \textbf{30} & 45.56 & 0.00 & 6.12 & 0.04 & 184 & 133.21 & 22.22 & Sgr/Car$^{1,2}$, W51$^{3}$ \\ \hline
        \textbf{31} & 47.39 & 0.00 & 3.21 & 0.01 & 210 & 124.56 & 16.31 & \\ \hline
        \textbf{32} & 47.44 & -6.43 & 6.69 & 0.74 & NA & 106.22 & NA & Sgr/Car$^{1}$ \\ \hline
        \textbf{33} & 49.52 & -4.35 & 9.89 & 0.18 & 86 & 114.51 & 8.31 & Per$^{1}$ \\ \hline
        \textbf{34} & 52.00 & -7.24 & 5.95 & 0.21 & 88 & 109.05 & 6.16 & \\ \hline
        \textbf{35} & 55.42 & -5.62 & 7.66 & 0.15 & 93 & 116.84 & 12.46 & Per$^{1}$ \\ \hline
        \textbf{36} & 55.76 & 0.00 & 4.55 & 0.06 & 135 & 110.78 & 6.34 & LoS$^{2}$ \\ \hline
        \textbf{37} & 56.62 & 12.46 & 3.47 & 0.18 & 68 & 108.05 & 2.00 & \\ \hline
        \textbf{38} & 57.02 & -2.39 & 9.01 & 0.30 & 67 & 114.13 & 4.56 & Per$^{1}$ \\ \hline
        \textbf{39} & 63.55 & 3.90 & 11.03 & 1.50 & 63 & 102.48 & 0.78 & \\ \hline
        \textbf{40} & 65.36 & 4.73 & 9.10 & 0.31 & 50 & 107.21 & 4.25 & \\ \hline
        \textbf{41} & 71.69 & 0.00 & 1.17 & 0.01 & 129 & 119.18 & 13.50 & Loc$^{1}$ \\ \hline
        \textbf{42} & 78.05 & 1.21 & 3.73 & 0.01 & 206 & 122.13 & 14.09 & Loc$^{1}$ \\ \hline
        \textbf{43} & 80.54 & 0.00 & 2.07 & 0.01 & 333 & 131.14 & 32.54 & Loc$^{1}$, Cygnus$^{3}$ \\ \hline
        \textbf{44} & 91.79 & 0.00 & 3.84 & 0.11 & 55 & 110.66 & 6.62 & Per$^{1}$ \\ \hline
        \textbf{45} & 122.28 & 0.00 & 1.12 & 0.03 & 36 & 101.30 & 0.90 & \\ \hline
        \textbf{46} & 133.83 & 0.00 & 3.47 & 0.06 & 100 & 105.46 & 4.94 & \\ \hline
        \textbf{47} & 141.93 & -11.13 & 3.88 & 0.08 & 46 & 110.86 & 6.14 & \\ \hline
        \textbf{48} & 148.17 & 23.72 & 1.86 & 0.03 & 61 & 113.41 & 7.91 & \\ \hline
        \textbf{49} & 149.04 & 40.61 & 1.15 & 0.02 & 44 & 110.70 & 7.28 & \\ \hline
        \textbf{50} & 177.52 & 0.00 & 4.30 & 0.02 & 148 & 186.97 & 58.58 & \\ \hline
        \textbf{51} & 180.81 & 3.79 & 5.67 & 0.18 & 55 & 106.82 & 5.41 & \\ \hline
        \textbf{52} & 192.30 & 0.00 & 2.14 & 0.02 & 82 & 110.70 & 5.50 & Per$^{1}$ \\ \hline
        \textbf{53} & 216.79 & -15.77 & 2.76 & 0.06 & 219 & 104.14 & 2.96 & \\ \hline
        \textbf{54} & 223.35 & 0.00 & 2.46 & 0.02 & 78 & 114.39 & 10.27 & Per$^{1}$ \\ \hline
        \textbf{55} & 269.05 & 0.00 & 2.14 & 0.01 & 285 & 151.51 & 31.18 & Vela$^{3}$ \\ \hline
        \textbf{56} & 281.73 & 0.00 & 4.06 & 0.10 & 104 & 107.18 & 3.75 & \\ \hline
        \textbf{57} & 284.02 & 0.00 & 5.71 & 0.05 & 219 & 124.63 & 15.20 & RCW 49$^{3}$ \\ \hline
        \textbf{58} & 291.12 & 0.00 & 4.72 & 1.35 & NA & 102.75 & NA & \\ \hline
        \textbf{59} & 297.85 & 0.00 & 3.32 & 0.04 & 133 & 135.06 & 23.29 & \\ \hline
        \textbf{60} & 304.08 & 10.35 & 4.17 & 1.12 & NA & 100.45 & NA & Sct/Cen$^{1}$ \\ \hline
        \textbf{61} & 327.62 & -4.62 & 9.31 & 0.79 & 69 & 105.84 & 4.23 & \\ \hline
        \textbf{62} & 342.06 & 0.00 & 4.36 & 0.03 & 249 & 139.16 & 26.83 & \\ \hline
        \textbf{63} & 343.83 & 0.00 & 2.21 & 0.02 & 126 & 120.40 & 11.59 & Sct/Cen$^{1}$, RCW 116B$^{3}$ \\ \hline
        \textbf{64} & 346.18 & 0.00 & 1.26 & 0.02 & 102 & 111.57 & 7.80 & Sgr/Car$^{1}$, NGC 6334$^{3}$ \\ \hline
        \textbf{65} & 346.97 & 0.00 & 6.21 & 0.79 & 69 & 101.60 & 1.32 & Sgr/Car$^{1}$ \\ \hline
        \textbf{66} & 348.32 & 2.35 & 9.15 & 0.64 & 58 & 109.20 & 1.51 & \\ \hline
        \textbf{67} & 348.76 & -4.40 & 9.78 & 0.11 & 117 & 125.47 & 17.94 & \\ \hline
        \textbf{68} & 349.55 & 4.28 & 10.06 & 0.13 & 99 & 120.64 & 9.04 & \\ \hline
        \textbf{69} & 349.55 & -4.86 & 8.86 & 0.09 & 73 & 122.07 & 14.83 & \\ \hline
        \textbf{70} & 351.15 & 5.07 & 8.48 & 0.22 & 92 & 111.12 & 9.90 & \\ \hline
        \textbf{71} & 352.25 & -3.82 & 5.37 & 0.05 & 113 & 109.99 & 6.97 & \\ \hline
        \textbf{72} & 352.82 & 6.68 & 6.44 & 0.25 & 55 & 108.92 & 5.02 & \\ \hline
        \textbf{73} & 353.48 & -4.94 & 7.63 & 0.15 & 155 & 113.17 & 10.45 & \\ \hline
        \textbf{74} & 354.05 & -2.34 & 9.20 & 0.10 & 107 & 132.36 & 15.22 & \\ \hline
        \textbf{75} & 354.20 & 0.21 & 4.51 & 0.01 & 201 & 200.24 & 68.41 & \\ \hline
        \textbf{76} & 354.41 & 5.72 & 7.52 & 0.11 & 109 & 121.47 & 13.62 & \\ \hline
        \textbf{77} & 356.05 & -4.70 & 6.74 & 0.04 & 171 & 133.86 & 18.02 & \\ \hline
        \textbf{78} & 356.37 & -13.37 & 3.24 & 0.33 & 69 & 101.74 & 1.86 & \\ \hline
        \textbf{79} & 357.02 & -8.87 & 4.86 & 0.35 & 4 & 103.64 & 4.97 & \\ \hline
        \textbf{80} & 357.40 & 2.16 & 9.94 & 0.20 & 79 & 129.49 & 17.01 & \\ \hline
        \textbf{81} & 357.54 & 0.00 & 7.07 & 0.03 & 173 & 171.91 & 52.88 & \\ \hline
        \textbf{82} & 358.07 & -0.83 & 8.18 & 0.03 & 134 & 143.10 & 27.47 & \\ \hline
        \textbf{83} & 358.47 & -2.39 & 8.99 & 0.04 & 137 & 167.76 & 42.20 & \\ \hline
        \textbf{84} & 359.94 & -5.24 & 8.21 & 0.05 & 75 & 126.60 & 18.88 & \\ \hline
\end{longtable}
\end{center}

\clearpage
\twocolumn

\newpage

\begin{appendix}
\let\part=\chapter\appendix

\section{Predicted uncertainties}
\label{sec:appendix}
\begin{figure}
    \centering
    \includegraphics[width=0.49\textwidth]{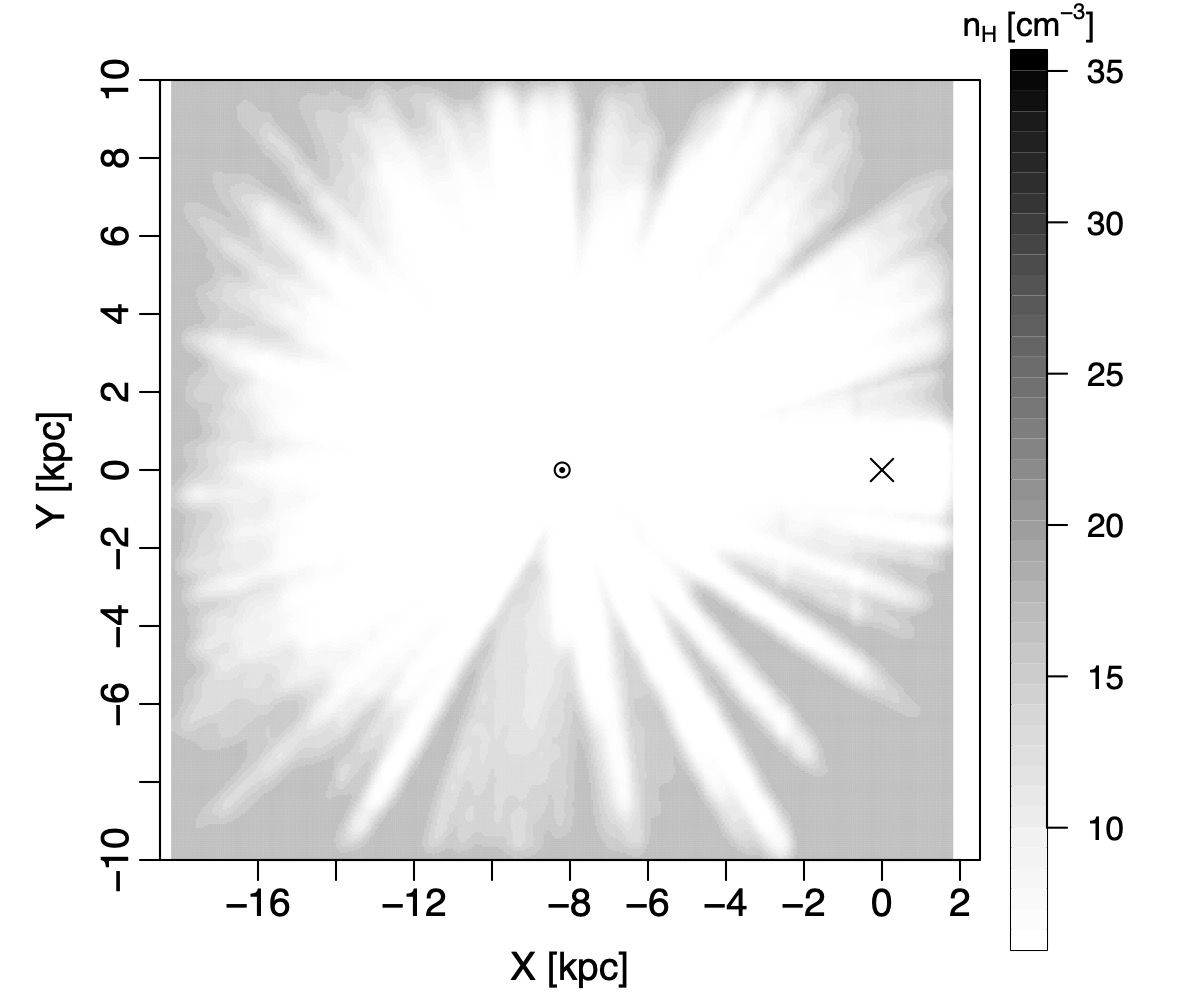}
    \caption{Predicted uncertainty for our 3D dust map. The Galactic Centre is at (0,0), marked by a cross.}
    \label{fig:uncertainties}
\end{figure}

Figure \ref{fig:uncertainties} shows our predicted uncertainties on the plane of the sky. A lack of input data from APOGEE causes the radial pattern. The fractional uncertainties in the dark areas are above 50\% and are the regions we marked in Fig. \ref{fig:3D_all}, left panel. Our selected higher densities are within low-uncertainty regions and have fractional uncertainties as low as 5\%.

\end{appendix}

\end{document}